\documentclass[11pt]{article}
\pdfoutput=1

\usepackage{jheppub}

\usepackage{amsmath, amssymb} 
\usepackage{mathtools}
\usepackage{esint}
\usepackage{bm}
\usepackage{dsfont}
\usepackage{braket}
\usepackage{eqparbox}
\usepackage{enumitem}
\hypersetup{
	colorlinks,
	urlcolor=Maroon,
	linkcolor=Maroon,
	citecolor=Maroon
	}

\numberwithin{equation}{section}

\usepackage[dvipsnames,table]{xcolor}
\usepackage{soul} 
\usepackage{calc}
\usepackage{subcaption}
\usepackage{tikz}
\usepackage{tkz-euclide} 
\usetikzlibrary{arrows.meta}
\tikzset{hidden/.style = {thick, dashed}}
\usepackage{scalefnt}
\usepackage[font=small,labelfont=bf]{caption}
\usepackage{graphicx}

\usepackage[numbers, sort&compress]{natbib}

\usepackage{comment} 
\usepackage[colorinlistoftodos]{todonotes}

\usepackage{hyperref}
\hypersetup{colorlinks=true, linktoc=page, linkcolor=Maroon, citecolor=Maroon}

\newcommand{\<}{\langle}
\renewcommand{\>}{\rangle}
\newcommand{\Tr}{\text{Tr}}

\newcommand{\be}{\begin{equation}}
\newcommand{\ee}{\end{equation}}
\newcommand{\bea}{\begin{eqnarray}}
\newcommand{\eea}{\end{eqnarray}}
\newcommand{\reef}[1]{(\ref{#1})}

\newcommand{\pa}{\partial}
\newcommand{\eps}{\epsilon}
\newcommand{\lra}{\leftrightarrow}
\def\ab#1{\langle#1\rangle} 
\def\sb#1{[#1]} 
\usetikzlibrary{decorations}

\usepackage{multirow}

\title{Gravitational Effective Theories with Maximal Supersymmetry and a Peculiar Parity}

\date{\today}

\author[a]{Justin Berman,}
\author[b]{Simon Caron-Huot,}
\author[a]{Aditi V. Chandra,}
\author[a]{Henriette Elvang,}
\author[c]{Aidan Herderschee,}
\author[a]{Loki L. Lin,}
\author[a]{Roger Morales}

\affiliation[a]{Leinweber Institute for Theoretical Physics, Randall Laboratory of Physics, University of Michigan, 450 Church St, Ann Arbor, MI 48109-1040, USA}
\affiliation[b]{Department of Physics, McGill University, 3600 Rue University, Montréal, H3A 2T8, QC Canada}
\affiliation[c]{Institute for Advanced Study, Einstein Drive, Princeton, NJ 08540, USA}

\emailAdd{jdhb@umich.edu}
\emailAdd{schuot@physics.mcgill.ca}
\emailAdd{adch@umich.edu}
\emailAdd{elvang@umich.edu}
\emailAdd{aidanh@ias.edu}
\emailAdd{lokilin@umich.edu}
\emailAdd{rmespasa@umich.edu}

\preprint{LITP-26-13}

\abstract{
We study the space of four-dimensional ultraviolet completions for $\mathcal{N}=8$ supergravity that are described at low energies by weakly-coupled effective field theories (EFTs) with maximal supersymmetry and $\mathrm{SU}(4)\times\mathrm{SU}(4)$ R-symmetry. We show that tree-level factorization of the 4-, 5-, and 6-point EFT scattering amplitudes, together with a certain ``peculiar parity'' condition, leads to nonlinear constraints on the 4-point Wilson coefficients. This peculiar parity is a property that can only be imposed on a subset of scalar amplitudes. Combining the nonlinear constraints with positivity, we find that the allowed region of 4-point Wilson coefficients is reduced to a non-convex domain with two sharp corners: one being the closed superstring Virasoro--Shapiro amplitude, the other an infinite spin tower amplitude exchanging states of every spin at the same mass. We show both numerically and analytically that requiring a finite number of states near the first mass level leaves only the Virasoro--Shapiro amplitude. 
}

\begin{document}

\maketitle

\addtocontents{toc}{\protect\setcounter{tocdepth}{2}}

\section{Introduction}
\label{sec:intro}

String theory is a remarkably rigid framework for quantum gravity, raising the possibility that a small set of general principles may select it as the unique ultraviolet (UV) completion of gravity. Establishing such a result in full generality remains a fundamental but formidable problem. Motivated by a growing body of bottom-up work \cite{Huang:2020nqy,Chiang:2023quf,Arkani-Hamed:2023jwn,
Cheung:2024uhn,Berman:2024wyt,Cheung:2024obl,Bhat:2024agd,Berman:2025owb,
Cheung:2025tbr,Eckner:2025kve,Elvang:2026pmc,Basile:2026gnd,Wan:2026pjq,
Xu:2026kix}, we instead ask a more modest question: how strongly can physical principles, supplemented
by possible additional assumptions, constrain the scattering amplitudes of a particular class of gravitational effective field theories (EFTs)?

We work in four dimensions, both because this is the dimensionality of our universe and because the relevant amplitudes remain tractable while retaining nontrivial kinematic structure. Our starting point is $\mathcal{N}=8$ supergravity (SUGRA) \cite{Cremmer:1978ds,Cremmer:1979up}, the maximally supersymmetric theory of gravity, supplemented by a general tower of higher-derivative interactions with a priori unconstrained coefficients, the so-called Wilson coefficients. We assume that the theory admits a weak-coupling limit in which the amplitudes exhibit tree-level factorization, with no contributions from loops of massless states, and restrict ourselves to these tree-level amplitudes. The closed superstring provides one possible UV completion of such an EFT, and the central question we ask is whether a minimal set of assumptions uniquely selects the closed string completion.

Along with $\mathcal{N}=8$ supersymmetry (SUSY), the leading-order SUGRA theory has an SU$(8)$ R-symmetry. 
Assuming the same R-symmetry to hold for the higher-derivative corrections does not allow for couplings such as $e^{w \varphi} R^n$ with the dilaton $\varphi$, where $R$ is a Riemann tensor and $w$ some number. To be able to study the broader class of graviton-dilaton EFTs with $\mathcal{N}=8$ SUSY, we relax the R-symmetry to $\text{SU}(4) \times\text{SU}(4)$. Two real singlets then combine into a complex axio-dilaton and the remaining 68 real scalars from the SUGRA supermultiplet decompose into two sets transforming in the $(\mathbf{4},\overline{\mathbf{4}})$ (or its conjugate) or the $(\mathbf{6},\mathbf{6})$ irreducible representations (irreps) of $\text{SU}(4) \times\text{SU}(4)$.

Discrete symmetries of the effective theory are also important in its definition. As we review below, with $\text{SU}(4) \times\text{SU}(4)$ R-symmetry, the largest possible discrete symmetry is $\mathbb{Z}_2\times \mathbb{Z}_2$, where the first factor interchanges the two $\text{SU}(4)$'s and the second factor is analogous to $CP$ in the Standard Model. However, we do not expect these symmetries to significantly affect our conclusions.

A more significant (and subtle) condition is \emph{peculiar parity}: we impose parity invariance for amplitudes whose external states are scalars in the $(\mathbf{6},\mathbf{6})$ irrep of $\mathrm{SU}(4)\times\mathrm{SU}(4)$. In other words, we demand that contractions with the Levi–Civita tensor $\epsilon_{\mu\nu\rho\sigma}$ are absent in such amplitudes. We call this condition peculiar to distinguish it from the previously mentioned symmetries.
What makes it peculiar is that its action cannot be defined on all states: owing to the chiral nature of R-charges and the SUSY algebra, charge conjugation $C$ and spacetime parity $P$ do not separately admit invariant actions on the fermions nor on all of the bosons.  Therefore, this property is unavoidably broken at loop level and no symmetry postulate can forbid the odd-parity tensor $\epsilon_{\mu\nu\rho\sigma}$ from appearing in $C$-odd combinations of bosonic scattering amplitudes on which $C$ acts.

This peculiar parity is somewhat analogous to $P$ in the Standard Model. Although $P$ (unlike $CP$) is violated by the chiral gauge couplings in processes involving external fermions, as well as by fermion loops in purely bosonic amplitudes, tree-level amplitudes of bosons preserve it. To the extent that the unavoidable breaking effects are subdominant, assuming such an approximate symmetry can restrict beyond-the-Standard Model scenarios. In the same sense, asking for specific bosonic amplitudes in an $\mathcal{N}=8$ effective theory to preserve $P$ at tree-level can narrow down possible UV scenarios.

The surprising constraining power of a peculiar parity in the S-matrix bootstrap was first observed in~\cite{Elvang:2026pmc} for $\mathcal{N}=4$ super Yang–Mills (SYM) EFTs. The analysis of \cite{Elvang:2026pmc} imposed maximal $\mathcal{N}=4$ SUSY, unbroken $\text{SU}(4)$ R-symmetry, planarity, and tree-level factorization, and asked in addition that the scalar amplitudes respect parity, i.e.~that they are free of Levi–Civita contractions of the external momenta. With such parity-odd contact terms excluded, SUSY and factorization impose an infinite set of \emph{nonlinear} relations among the 4-point Wilson coefficients. These relations leave undetermined only the coefficient of $\Tr F^4$ together with a single new coefficient for every second order in the derivative expansion starting at order $\Tr D^2 F^4$, i.e.~only one free parameter for each set of operators of the form $\Tr D^{2(2m+1)} F^4$ with $m=0,1,2,\dots$. The nonlinear constraints were also shown to imply the open-string monodromy relations~\cite{Plahte:1970wy,Stieberger:2009hq,Bjerrum-Bohr:2009ulz,Bjerrum-Bohr:2010mia,Bjerrum-Bohr:2010pnr} order by order in the derivative expansion, and to numerically select the Veneziano amplitude. Monodromy together with positivity was in turn recently proven to single out the Veneziano amplitude to all orders~\cite{Wan:2026pjq}.\footnote{Earlier work~\cite{Huang:2020nqy,Berman:2023jys,Chiang:2023quf} also found numerical evidence for this.} Maximal SUSY, R-symmetry, planarity, tree-level factorization, peculiar parity, and positivity therefore isolate the open string as the unique consistent UV completion of $\mathcal{N}=4$ SYM at tree level. With parity-odd terms allowed, however, the same requirements do not yield nonlinear relations: peculiar parity appears to be the essential property that singles out the Veneziano amplitude.

In this paper, we analyze the $\mathcal{N}=8$ SUSY constraints for tree-level amplitudes, considering both SU$(8)$ and $\text{SU}(4) \times\text{SU}(4)$ R-symmetry. At 4-point, the amplitudes are insensitive to the R-symmetry breaking, and the $\mathcal{N}=8$ SUSY Ward identities imply that all 4-point component amplitudes are determined by a single function of the Mandelstam invariants~\cite{Grisaru:1976vm,Grisaru:1977px,Bianchi:2008pu,Elvang:2013cua}. Choosing the graviton component amplitude $M_4(--+\,+)$ as a representative, we have\footnote{We take all external momenta $p_i$ as outgoing and the Mandelstam variables as $s_{i_1\cdots i_n}=(p_{i_1}+\cdots+p_{i_n})^2$. At 4 points, we further define $s\equiv s_{12}=\langle12\rangle[12]$, $t\equiv s_{13}=\langle13\rangle[13]$, and $u\equiv s_{14}=\langle14\rangle[14]$, with $s+t+u=0$.}
\be
\label{M4intro}
M_4(--+\,+)=\langle12\rangle^4[34]^4 F(s,t,u)\,.
\ee
Crossing symmetry requires $F(s,t,u)$ to be fully permutation symmetric, and its low-energy expansion encodes the higher-derivative corrections to the leading two-derivative amplitude. Concretely, it takes the form
\begin{align}\label{F4pt}
    F(s,t,u)= &\, \frac{\kappa^2}{stu} + g_{0} + g_{2} \, (s^2+t^2+u^2) + g_{3} \, stu + g_{4} \, (s^2+t^2+u^2)^2 + g_{5} \, stu \,(s^2+t^2+u^2) \nonumber \\ 
    & + g_6 \, (s^2+t^2+u^2)^3 + g_6' (stu)^2 + \dots  \,,
\end{align}
where $\kappa^2 = 8\pi G_N$ with $G_N$ being Newton's constant. The first term, $\kappa^2/stu$, is the two-derivative SUGRA amplitude generated by a graviton exchange in the three channels, while the analytic remainder is a power series whose coefficients are the Wilson coefficients of the higher-derivative operators. These are in one-to-one correspondence with local operators $D^{2m}R^4$ in the effective action: $g_0$ is the coefficient of $R^4$, $g_2$ that of $D^4R^4$, $g_3$ that of $D^6R^4$, and so on, possibly dressed with a dilaton coupling that does not affect 4-point scattering. 

To constrain these coefficients appearing at 4 points and beyond, we construct the 6-point NMHV superamplitude in the EFT and subject it to the requirements of $\mathcal{N}=8$ SUSY and consistent factorization. We impose the peculiar parity condition on the amplitude 
\begin{equation}
  \mathbb{Z}_{1,1} \equiv M_6(z^{1256} z^{1357} z^{1458} z^{2367} z^{3478} z^{2468})\,,
\end{equation}
with all 6 external states from the $(\mathbf{6},\mathbf{6})$ scalar sector. Practically, this amounts to requiring it to be free of parity-odd Levi–Civita contractions of the external momenta.

The construction of the 6-point amplitudes is highly sensitive to the R-symmetry. 
{\bf With unbroken  SU(8) R-symmetry},
we find that when combined with positivity constraints: 
\begin{itemize}
\item[$\vcenter{\hbox{\tiny$\bullet$}}$] imposing peculiar parity sets all higher-derivative corrections to zero;
\item[$\vcenter{\hbox{\tiny$\bullet$}}$] extending the vanishing scalar soft limits known from the leading-order $\mathcal{N}=8$ SUGRA theory is only possible when all higher-derivative corrections are zero.
\end{itemize}
This shows that the only unitary EFT at tree-level with either of these two properties is the pure $\mathcal{N}=8$ SUGRA theory.

In contrast, a much richer structure is allowed when the {\bf R-symmetry is relaxed to $\text{SU(4)} \times\text{SU(4)}$}:
\begin{itemize}
    \item[$\vcenter{\hbox{\tiny$\bullet$}}$] peculiar parity implies novel nonlinear constraints among the 4-point Wilson coefficients. 
\end{itemize}
At the lowest orders, these nonlinear relations are 
\begin{equation}
\label{eq:nonlinear_constraints_intro}
    \kappa^2 g_3 = \frac{1}{2}g_0^2\,, \quad~ \kappa^2 g_5 = g_2 g_0\,, \quad~ \kappa^4g_6' = \frac{8}{3}\kappa^4 g_6 + \frac{1}{6}g_0^3\,, \quad~ \kappa^2g_7 = g_4g_0 + \frac{1}{2}g_2^2\,, \ \dots.
\end{equation}
We have continued this analysis up to $g_9$ and $g_9'$, and found that all Wilson coefficients are fixed in terms of the coefficients $g_{2m}$, i.e. $g_0$, $g_2$, $g_4,\dots$. (Note that these are the coefficients that survive in \eqref{F4pt} when $u=0$.)

One might ask whether further relations exist beyond these, ones that would also fix the remaining coefficients $g_{2m}$. The case of closed superstring theory suggests that additional algebraic relations do not exist. The 4-graviton Virasoro–Shapiro amplitude in closed superstring theory~\cite{Virasoro:1969me,Shapiro:1970gy} is given by
\begin{equation}
\label{eq:Virasoro-Shapiro_intro}
   M_4^{\text{str}}(--+\,+) = \langle12 \rangle^4 [34]^4 \, \frac{\kappa^2}{stu} \,\frac{\Gamma(1- \alpha's) \Gamma(1- \alpha't)\Gamma(1-\alpha' u)}{\Gamma(1+\alpha's) \Gamma(1+\alpha't)\Gamma(1+\alpha' u)} \,.
\end{equation}
Its low-energy $\alpha'$-expansion takes the form of~\eqref{F4pt} with Wilson coefficients
\begin{equation}
\label{eq:Wilson_coeffs_VS_intro}
\begin{split}
    g_0 =&\, 2 \kappa^2 \alpha'^3 \zeta_3\,, \qquad~~\, g_2=\kappa^2\alpha'^5 \zeta_5\,, \qquad~~\, g_3 = 2\kappa^2\alpha'^6 \zeta_3^2\,, \qquad~~\, g_4 = \frac{1}{2} \kappa^2\alpha'^7 \zeta_7\,, \\[1.5mm]
    g_5 =&\,2 \kappa^2\alpha'^8 \zeta_3 \zeta_5\,, \qquad\hspace{-0.03cm} g_6 = \frac{1}{4} \kappa^2 \alpha'^9 \zeta_9\,, \qquad\hspace{0.06cm} g_6' = \frac{4}{3} \kappa^2 \alpha'^9 \zeta_3^3 + \frac{2}{3} \kappa^2 \alpha'^9 \zeta_9\,, \ \ \dots\ .
\end{split}
\end{equation}
As can be seen, the Wilson coefficients are expressed in terms of the odd Riemann zeta values, where $\zeta_b = \sum_{n=1}^\infty 1/n^b$, and they satisfy the nonlinear constraints from~\eqref{eq:nonlinear_constraints_intro}. Importantly, we notice that the Wilson coefficients $g_{2m}$ left unfixed by the nonlinear constraints precisely align with the first occurrence of each $\zeta_\text{odd}$, since $g_{2m} = 1/2^{m-1} \kappa^2 \alpha'^{2m+3} \zeta_{2m+3}$ for $m=0$, $1$, $2,\dots$. Because the $\zeta_\text{odd}$ are conjecturally algebraically independent, no further algebraic constraints can be expected among the Wilson coefficients $g_{2m}$.

The nonlinear relations \reef{eq:nonlinear_constraints_intro} are derived bottom-up and must hold for any UV completion consistent with the stated assumptions. On their own, however, they do not fix the 4-point amplitude, but parameterize it in terms of the $g_{2m}$ coefficients. To constrain it further, we turn to positivity. Not all values of the Wilson coefficients are compatible with a unitary and causal UV completion~\cite{Adams:2006sv} and the modern S-matrix bootstrap sharpens this into quantitative bounds~\cite{Caron-Huot:2016icg,Caron-Huot:2020cmc,Tolley:2020gtv,Bellazzini:2020cot,Arkani-Hamed:2020blm,Albert:2022oes}. Assuming unitarity, analyticity, and that the scalar amplitude $M_4(zz\bar{z}\bar{z})/s^3$ vanishes at large $|s|$ for fixed $u<0$, a dispersion relation expresses each coefficient $g_{2m}$ as a positive sum over the spins exchanged above the mass gap. Together with the crossing (null) constraints, this casts the bounds on the Wilson coefficients as a semi-definite optimization problem, into which we incorporate the nonlinear relations~\eqref{eq:nonlinear_constraints_intro}.

In the plane of the ratios $(g_2/g_0,\, g_3/g_0)$, the nonlinear relations reduce the allowed region to a non-convex shape with two corners, which get sharper as more nonlinear relations and null constraints are imposed. The first corner converges to the Virasoro–Shapiro amplitude (\ref{eq:Virasoro-Shapiro_intro}). The second is a particular version of the Infinite Spin Tower (IST),
\begin{align}
\label{eq:IST_intro_2}
      M_4^{\text{IST}}(--+\,+) = \<12\>^4 [34]^4 \, \frac{\kappa^2}{stu}\, \frac{(m^2+s)(m^2+t)(m^2+u)}{(m^2-s)(m^2-t)(m^2-u)}\,,
\end{align}
which exchanges states of every spin $J=0$, $2$, $4$, $\dots$ at the same mass $m$. The coefficients of the low-energy expansion of \reef{eq:IST_intro_2} satisfy all of the nonlinear relations~\eqref{eq:nonlinear_constraints_intro}.
We posit that the boundary of the allowed region in the $(g_2/g_0,\, g_3/g_0)$-plane is determined by mass-scaling curves for the corner theories, together with a line between the corners corresponding to a novel family of modified IST amplitudes that we construct explicitly.

Requiring that only finitely many spins appear at the mass gap removes all IST amplitudes, and the allowed region bifurcates into a small island around the Virasoro–Shapiro amplitude and a smaller, scaled down non-convex region. We illustrate this numerically and then prove it analytically. Specifically, the nonlinear relations resum the 4-point amplitude into an exponential form whose spectrum, once only finitely many spins are allowed at the gap, is forced to be the discrete tower of the closed string. In this way, 
the Virasoro–Shapiro amplitude arises as the unique solution to the supersymmetric EFT bootstrap. 

Following this summary of the results, we briefly comment on how the symmetries we impose arise in string theory. In the full quantum theory of closed superstrings, R-symmetry is known to be broken to a discrete duality group \cite{Hull:1994ys}.
Nonetheless, $\text{SU}(4) \times\text{SU}(4)$ holds as a symmetry of the tree-level scattering amplitudes, as is made manifest for example by the KLT double-copy formula \cite{Kawai:1985xq}, which expresses closed string tree-level amplitudes as sums of products of open string amplitudes. The $\mathcal{N}=4$ SUSY of each factor combine into an exact $\mathcal{N}=8$ SUSY, while the $\text{SU}(4) \times\text{SU}(4)$ R-symmetry only enhances to SU$(8)$ in the leading-order SUGRA limit; the discrete symmetry which swaps the two factors is realized by T-duality.
As mentioned earlier, from the low-energy perspective, $\text{SU}(4) \times\text{SU}(4)$ is the minimal amount of breaking that enables a dilaton to be singled out.

Peculiar parity also has an origin in string theory. One finds that tree-level closed strings automatically inherit peculiar parity via the KLT double-copy of the open string, so it suffices to consider the latter. Technically, parity of tree-level bosonic amplitudes in the open superstring follows from the fact that the vertex operators for bosons, which reside in the so-called NS sector, can be expressed using only worldsheet fields that carry spacetime vector indices and whose correlators are parity symmetric \cite{Elvang:2026pmc}. In contrast, fermion vertex operators reside in the so-called R sector, where worldsheet fermion zero modes break parity. Peculiar parity in string theory can thus be attributed to a discrete conservation law on the worldsheet that decouples the NS sector from the R sector at genus 0~\cite{Friedan:1985ge,Witten:2012ga}.

From the spacetime perspective, the fact that fermion loops unavoidably violate peculiar parity indicates that it is at best an approximate symmetry (of the subset of amplitudes on which it is defined). 
What is remarkable in string theory is that this approximate symmetry is not confined to the low-energy regime: its breaking remains at the minimal level required by loops, even at the string scale. Since it entails the vanishing of infinitely many Wilson coefficients in the weak-coupling limit, one might say that peculiar parity \emph{emanates} from a property of string worldsheets, to borrow a terminology used for exact IR symmetries that emanate from distinct symmetries of short-distance lattice Hamiltonians \cite{Seiberg:2023cdc}. This analogy is imperfect, however, because unlike a conventional symmetry, it does not seem possible to define peculiar parity on all states.
 
\subsection*{Outline of the Paper}

The remainder of this paper is organized as follows. Section~\ref{sec:symmetries} briefly describes the spectrum and symmetries of $\mathcal{N}=8$ SUGRA and superstring theory compactified on a six-torus, including the breaking of R-symmetry from SU$(8)$ to $\text{SU}(4) \times\text{SU}(4)$, the discrete symmetries, and peculiar parity. In Section~\ref{sec:Superamps_SUGRA}, we introduce the on-shell superspace formalism and construct the 3-, 4- and 5-point superamplitudes in $\mathcal{N}=8$ SUSY EFTs. In Section~\ref{sec:Rsym_SU4xSU4}, we construct the 6-point NMHV superamplitude with $\text{SU}(4) \times\text{SU}(4)$ R-symmetry; we  examine unbroken SU$(8)$  as a special case. Peculiar parity is then imposed in Section~\ref{sec:PP}, where we derive the resulting nonlinear constraints among the 4-point Wilson coefficients. We also study the soft limits for the scalars, and test the nonlinear constraints on various amplitudes, including the Virasoro–Shapiro amplitude and the Infinite Spin Tower. Section~\ref{sec:positivity} contains the numerical analysis with the positivity bounds as well as the analytic bootstrap of the Virasoro–Shapiro amplitude. We conclude in Section~\ref{sec:conclusions} with a discussion of future directions.

This paper also includes five appendices. In Appendix~\ref{app:5pt_analysis}, we present the details for the calculation of the 5-point MHV superamplitude. In Appendix~\ref{app:6ptHelSect}, we collect the remaining 6-point helicity sectors, namely the MHV, N$^{(1,0)}$MHV and N$^{(2,0)}$MHV amplitudes. In Appendix~\ref{app:counting}, we explain the O$(4)$-character method used to count the $S_2\times S_4$-symmetric polynomials in our ans\"atze. In Appendix~\ref{app:Rsym_SU8}, we develop the 6-point NMHV superamplitude with unbroken SU$(8)$ R-symmetry. Lastly, in Appendix~\ref{app:comparison_string}, we review the KLT formulas for the 4-, 5- and 6-point closed string amplitudes, which are used as consistency checks for our EFT amplitudes.
\\
\\
\noindent {\bf Note added:} When this work was in the final stages of being written up, another paper \cite{Xu:2026kix} appeared with partial overlap. The author of \cite{Xu:2026kix} used the result of the $\mathcal{N}=4$ EFT analysis in \cite{Elvang:2026pmc} and assumed the string theory KLT double copy to motivate an exponential ansatz for the gravitational 4-point amplitude, which was used to deduce nonlinear constraints among the 4-point Wilson coefficients. In contrast, we derive the nonlinear constraints from maximal SUSY, tree-level factorization, and peculiar parity, and use them to deduce the exponentiation of the gravitational amplitude.

\section{Spectrum and Symmetries}
\label{sec:symmetries}

In this section, we review the spectrum, R-symmetry and discrete symmetries of $\mathcal{N}=8$ SUGRA with and without higher-derivative corrections.

\subsection{Spectrum and Symmetries of \texorpdfstring{$\mathcal{N}=8$}{N=8} SUGRA and Beyond}
\label{sec:supermultiplet_R-sym}

The 4d self-dual on-shell massless $\mathcal{N}=8$ supermultiplet consists of $2^{\mathcal{N}}=256$ states that  transform in the fully antisymmetric irreducible representations (irreps) of the SU$(8)$ R-symmetry. Using SU$(8)$  indices $a,b,\ldots=1,2,\dots,8$, the on-shell states are
\begin{itemize}
    \item[$\vcenter{\hbox{\tiny$\bullet$}}$] The positive and negative helicity gravitons, $h^+$ and $h^-=h^{12345678}$, which are SU$(8)$ singlets.
    \item[$\vcenter{\hbox{\tiny$\bullet$}}$] 8 pairs of positive and negative helicity gravitino states, $\psi^a$ and $\psi^{abcdefg}$, transforming in the fundamental and anti-fundamental of SU$(8)$, corresponding to $\mathbf{8} \oplus \overline{\mathbf{8}}$.
    \item[$\vcenter{\hbox{\tiny$\bullet$}}$] 28 pairs of positive and negative helicity gravi-photon states, $v^{ab}$ and $v^{abcdef}$, transforming in the 2- and 6-index antisymmetric irreps of SU$(8)$, i.e. $\mathbf{28} \oplus \overline{\mathbf{28}}$. 
    \item[$\vcenter{\hbox{\tiny$\bullet$}}$] 56 pairs of positive and negative helicity gravi-photino states $\chi^{abc}$ and $\chi^{abcde}$, transforming in the $\mathbf{56} \oplus \overline{\mathbf{56}}$ of SU$(8)$.
    \item[$\vcenter{\hbox{\tiny$\bullet$}}$] 70 real scalars $\phi^{abcd}$ that sit in the $\mathbf{70}$ of SU$(8)$, which can be combined into 35 complex scalars $z^{abcd}$ satisfying $\bar{z}_{abcd} = (z^{abcd})^*
 = \frac{1}{4!}\eps_{abcdefgh}
 z^{efgh}$.
\end{itemize}

Requiring full SU$(8)$ R-symmetry for the $\mathcal{N}=8$ EFT
restricts the bosonic operators to be of the schematic form $R^n$, $D^{2k} R^n$, $z \bar{z} R^n$, etc.,~where the $R$ are Riemann tensors, $D$ denotes covariant derivatives, and the $z$ are scalars in the massless SUGRA supermultiplet. 
In particular, SU$(8)$ excludes the possibility of dilatonic couplings such as $e^{w \varphi} R^n$, with $\varphi$ being the dilaton and $w$ some number, because none of the 70 scalars are SU$(8)$ singlets. The minimal R-symmetry breaking that allows us to study the broader class of graviton-dilaton EFTs is replacing SU$(8)$ with $\text{SU}(4) \times\text{SU}(4)$. Then, two real singlets combine into a complex axio-dilaton and the remaining 68 real scalars decompose into two sets: one transforming in the $(\mathbf{4},\overline{\mathbf{4}})$ (or its conjugate) irrep of $\text{SU}(4) \times\text{SU}(4)$ and one in the $(\mathbf{6},\mathbf{6})$. 
Specifically, splitting the R-indices into two SU$(4)$ groups,  $\{1,2,3,4\}$ and $\{5,6,7,8\}$, 
we have~\cite{Elvang:2010kc}:
\begin{enumerate}
\item $z^{1234}$ and its complex conjugate $z^{5678}$, which are singlets of $\text{SU}(4)\times\text{SU}(4)$. The real and imaginary parts can be identified as the dilaton $\varphi=\frac{1}{2}(z^{1234}+z^{5678})$ and the 4d axion $a=\frac{1}{2i}(z^{1234}-z^{5678})$.

\item 32 real scalars that sit in the fundamental of one SU$(4)$ and in the anti-fundamental of the other, i.e.~the $(\mathbf{4},\overline{\mathbf{4}}) \oplus (\overline{\mathbf{4}},\mathbf{4})$ irrep of $\text{SU}(4)\times\text{SU}(4)$. They can be combined into 16 complex scalars, e.g. $z^{123|5}$ and its conjugate $z^{4|678}$.

\item 36 real scalars that transform in the 2-index antisymmetric irrep of each of the SU$(4)$ factors, i.e.~the $(\mathbf{6},\mathbf{6})$ irrep of $\text{SU}(4)\times\text{SU}(4)$. They can be organized into 18 complex scalars, such as $z^{12|56}$.
    
\end{enumerate}

In the case of the low-energy EFT of the closed string at tree-level, the $\alpha'$-corrections  explicitly break the global SU$(8)$ R-symmetry to $\text{SU}(4) \times\text{SU}(4)$. The $\text{SU}(4) \times\text{SU}(4)$ symmetry is manifest in the KLT relations~\cite{Kawai:1985xq} that express the closed string tree-level amplitudes in terms of products of open string amplitudes, each of which contribute an SU$(4)$ symmetry. 
In fact, in the context of the double copy, the dilaton and axion arise from the double-copy of opposite helicity gluons, the $(\mathbf{4},\overline{\mathbf{4}}) \oplus (\overline{\mathbf{4}},\mathbf{4})$ scalars are double-copies of opposite-helicity gluinos, and the $(\mathbf{6},\mathbf{6})$  scalars arise from the double-copy of two scalars in the $\mathcal{N}=4$ supermultiplet. 

Let us now turn to soft theorems.  
In the leading-order $\mathcal{N}=8$ SUGRA theory, the equations of motion exhibit a continuous global $E_{7(7)}$ symmetry, which is spontaneously broken to SU$(8)$~\cite{Cremmer:1978ds,Cremmer:1979up}. The 70 real scalars are Goldstone modes of this breaking and live in the coset $E_{7(7)}/$SU$(8)$. As a result, the soft theorems guarantee that the $\mathcal{N}=8$ SUGRA amplitudes vanish in the single-scalar soft limit. For the closed string tree-level amplitude, the compactification from 10d to 4d on the six-torus gives an O$(6,6)$ T-duality symmetry. We can think of the $\alpha'$-corrections as breaking the $E_{7(7)}$ explicitly to O$(6,6)$. The spontaneous breaking is then from O$(6,6)$ to $\text{SU}(4) \times\text{SU}(4)$; this involves $66-30=36$ broken generators, giving rise to 36 Goldstone modes, identified as the $(\mathbf{6},\mathbf{6})$ scalars.  Thus, the tree-level closed superstring amplitude has vanishing single-soft limits for this class of scalars.

\subsection{Additional Discrete Symmetries and Peculiar Parity}
\label{ssec:discrete}

We now discuss the possible discrete extensions of $\text{SU}(4) \times\text{SU}(4)$ R-symmetry. Since the SUSY and R-symmetry in this setup are the double-copy of those of $\mathcal{N}=4$ SYM, it is helpful to first discuss that theory.

\def\Qbar{\widetilde{Q}}

The $\mathcal{N}=4$ matter content is most easily described as the dimensional reduction of ten-dimensional $\mathcal{N}=1$ SYM \cite{Brink:1976bc}, where the supercharge $Q_\alpha$ and the gluino are 10d Majorana–Weyl spinors with 16 real components. Ten-dimensional parity $P_{10}$ is not a symmetry: it interchanges two distinct but equivalent versions of the theory in which the chirality of the gluino is reversed.

Upon dimensionally reducing from 10d to 4d on the six-torus, i.e.~on $\mathbb{R}^{3,1}\times T^6$, the 16 real supercharges organize into four complex Weyl spinors $Q^a$ and their conjugate $\Qbar_{a}$, where $a=1, \ldots, 4$ are SU(4) R-symmetry indices.
The transformation $P_{10}$ can be dimensionally reduced in two different ways: either as spatial reflection $P{:}\ (x_1,x_2,x_3)\mapsto (-x_1,-x_2,-x_3)$, or as reflection of a compactified direction $C{:}\ x_9\mapsto -x_9$.  Our labels here follow their 4d interpretation: $P$ is the spacetime parity which interchanges holomorphic and anti-holomorphic spinors, while $C$ is the charge conjugation which exchanges upper and lower R-symmetry indices (i.e.~the ${\bf 4}$ and $\overline{\bf 4}$ of SU$(4)$).

Like $P_{10}$, neither $C$ nor $P$ separately admit an invariant action on the $\mathcal{N}=4$ fermions. Their R-charges are chiral and concretely the Yukawa couplings and SUSY transformations could not be preserved by such an action. This is analogous to $C$ and $P$ which separately do not act on Standard Model fermions (although in the Standard Model the obstruction lies in the chiral nature of gauge interactions rather than with the SUSY algebra). Only the product $CP$ can be imposed as a bona fide symmetry of an $\mathcal{N}=4$ EFT. This symmetry can be exact in specific UV deformations of ${\mathcal N}=4$ SYM, such as toroidal compactifications of the type I superstring.

Although $C$ and $P$ do not separately act on fermions, they do act on bosonic states and it makes sense to ask whether amplitudes involving only external bosons are invariant under both $C$ and $P$.  This is the \emph{peculiar parity} discussed in the introduction.  While fermion loops clearly violate this property, there can exist situations where these can be neglected and the question becomes nontrivial.  There is strong evidence that the type I superstring is the only tree-level UV modification of $\mathcal{N}=4$ SYM which enjoys a peculiar parity in this sense \cite{Elvang:2026pmc}. 

Moving on to $\mathcal{N}=8$ SUGRA
EFTs with $\text{SU}(4) \times\text{SU}(4)$ R-symmetry, it is possible to impose an exact $\mathbb{Z}_2\times \mathbb{Z}_2$ symmetry. The first generator is $CP$, where $C$ now acts on both $\text{SU}(4)$'s, while the second generator $S$ swaps the two $\text{SU}(4)$'s.  This symmetry is realized in specific EFTs such as toroidal compactifications of the IIA superstring.\footnote{In its standard parity-symmetric formulation, the supercharges of IIA are 10d Majorana (not-Weyl) spinors and its symmetries are the double-copy of those of two opposite-chirality type I theories. The $\mathbb{Z}_2\times \mathbb{Z}_2$ is then generated by the two distinct ways of dimensionally reducing the 10d symmetry $P_{10}$.  These become our $CP$ and $S$ upon switching to a distinct but equivalent formulation by applying $C$ to one of the $\text{SU}(4)$'s.}

The \emph{peculiar parity} that we impose for $\mathcal{N}=8$ is more subtle: we impose $P$ on amplitudes whose external states are $(\mathbf{6},\mathbf{6})$ scalars, on which $P$ admits a well-defined action. It turns out that not only is tree-level closed string theory invariant under peculiar parity, but this property quantitatively isolates the Virasoro–Shapiro amplitude as we show in Section~\ref{sec:positivity}.

Note that if $CP$ is imposed, $P$ is equivalent to $C$, which complex-conjugates the indices of both SU$(4)$'s. Through the identification SU(4)${}\simeq {\rm Spin}(6)$, $C$ is interpreted as a 6d parity which reflects one of the 6 scalar directions. (In the standard parity-symmetric formulation of IIA, $x_9\mapsto -x_9$ is also a peculiar parity: it is not a symmetry unless combined with swapping the two double-copy factors.) Since the first object that can break this symmetry is the 6d Levi–Civita tensor, peculiar parity can only be nontrivially tested for amplitudes involving $n\geq 6$ scalars.

\section{Amplitudes in \texorpdfstring{$\mathcal{N}=8$}{N=8} SUGRA EFTs}
\label{sec:Superamps_SUGRA}

In this section, we introduce the on-shell superspace formalism needed to construct the superamplitudes, which we calculate at 3, 4 and 5 points. Based on the symmetry discussion from the previous section, we directly restrict to the case of $\text{SU}(4)\times\text{SU}(4)$ R-symmetry; we discuss the case with unbroken SU$(8)$ R-symmetry in Section~\ref{sec:wSU(8)}. Note that we work bottom up in effective theory assuming only these ingredients: we do not assume string theory nor the double copy.

\subsection{On-shell Superamplitudes}
\label{sec:On-shell_superamps}

The SU$(8)$-invariant $\mathcal{N}=8$ supermultiplet states from Section~\ref{sec:supermultiplet_R-sym} can be conveniently arranged into a single on-shell superfield $\Phi_i$ for each external particle  $i$~\cite{Elvang:2008na,Elvang:2013cua,Elvang_Huang_2015},
\begin{equation}
    \Phi_i = h_i^+ + \eta_{ia} \psi_i^a - \frac{1}{2} \eta_{ia} \eta_{ib} v_i^{ab} + \ldots + \eta_{i1} \eta_{i2} \eta_{i3} \eta_{i4} \eta_{i5} \eta_{i6} \eta_{i7} \eta_{i8} h_i^-\,,
\end{equation}
where the $\eta_{ia}$ are Grassmann variables. This way, we  collect $n$-point amplitudes with different SUSY-related external states into a single superamplitude,
\be
 \mathcal{M}_n(123\ldots n) \equiv \mathcal{M}_n(\Phi_1 \Phi_2 \Phi_3 \ldots \Phi_n)\,.
\ee

In general, the superamplitude is a sum of polynomials of the Grassmann variables, which define different helicity sectors. Restricting to $\text{SU}(4) \times\text{SU}(4)$ R-symmetry, the superamplitude decomposes into sectors of degree $4k$ and $4\tilde{k}$ in the two sets of Grassmann variables $\{ \eta_{i1},\eta_{i2},\eta_{i3},\eta_{i4}\}$ and $\{ \eta_{i5},\eta_{i6},\eta_{i7},\eta_{i8}\}$, associated with the respective SU$(4)$ factors. These sectors are denoted as N$^{(k-2,\tilde{k}-2)}$MHV. 
The N$^{(k-2,k-2)}$MHV sector is the same as the SU$(8)$-invariant N$^{k-2}$MHV sector, which includes the amplitude with $k$ negative and $n-k$ positive helicity gravitons. The N$^{(1,0)}$MHV sector contains amplitudes such as $M_n(z^{1234}--++)$, whereas N$^{(2,0)}$MHV includes 
$M_n(z^{1234}z^{1234}--++)$, and so on. The supercharges only change the degree of the Grassmann polynomial by one, so the different helicity sectors do not mix under SUSY.

A specific component amplitude is extracted from the superamplitude by taking the Grassmann derivatives corresponding to the desired external states, i.e.
\begin{equation}
    h_i^+ \to 1\,, \ \quad \psi_i^a \to \partial_i^a\,, \ \quad v_i^{ab} \to \partial_i^a \partial_i^b\,, \ \quad \chi_i^{abc} \to \partial_i^a \partial_i^b \partial_i^c\,, \quad \dots\,, \quad h_i^- \to \partial_i^1 \partial_i^2 \partial_i^3 \partial_i^4 \partial_i^5 \partial_i^6 \partial_i^7 \partial_i^8\,,
\end{equation}
where $\partial_i^a = \partial/\partial \eta_{ia}$.\footnote{We use the package~\cite{Grassmann} for the \texttt{Mathematica} implementation of Grassmann variables and their derivatives.} For instance, we have that
\begin{equation}
    M_n(z^{1256} z^{3478}-+\ldots +) = (\partial_1^1 \partial_1^2 \partial_1^5 \partial_1^6 ) (\partial_2^3 \partial_2^4 \partial_2^7 \partial_2^8 ) (\partial_3^1 \cdots \partial_3^8 )\, \mathcal{M}_n(123\ldots n)\Big|_{\eta_{ia}=0}\,.
\end{equation}

Supersymmetry relates amplitudes with different external states via the supersymmetric Ward identities~\cite{Grisaru:1976vm,Grisaru:1977px}. In the on-shell superspace formalism, the SUSY Ward identities are equivalent to the annihilation of the superamplitude 
\begin{equation}
\label{eq:supercharges_action}
    Q^a \mathcal{M}_n = 0\,, \qquad \quad \widetilde Q_a \mathcal{M}_n = 0\,,
\end{equation}
by the supercharges
\begin{equation} 
\label{eq:def_supercharges}
Q^a = \sum_{i=1}^n\,[i|\, \frac{\pa}{\pa\eta_{ia}}\,, \qquad\quad
\widetilde Q_a = \sum_{i=1}^n\, | i\>\, \eta_{ia} \,.
\end{equation} 
When $\mathcal{N} = 8$ SUGRA is deformed with higher-derivative operators, the amplitudes continue to be related by SUSY Ward identities, so the degrees of freedom in the full superamplitude are determined by the symmetries respected by the underlying theory.

The SUSY Ward identities were solved for both $\mathcal{N}=4$ and $\mathcal{N}=8$ SUSY  in~\cite{Elvang:2009wd,Elvang:2010xn}, where it was shown that \reef{eq:supercharges_action} is solved for a superamplitude built from two basic building blocks, 
\begin{equation}
\label{eq:deltaQ}
  \delta^{(2\mathcal{N})}\big( \widetilde{Q}\big)
  \equiv 
  \frac{1}{2^{\mathcal{N}}}
  \prod_{a=1}^{\mathcal{N}} \sum_{1\leq i<j\leq n} \<ij\> \eta_{ia}\eta_{ja}\,,
\end{equation}
as well as
\begin{equation}
\label{eq:mijks}
    m_{ijk,a} \equiv [ij] \eta_{ka} +[jk] \eta_{ia} +[ki] \eta_{ja}\,.
\end{equation} 
The Grassmann delta function \reef{eq:deltaQ} manifestly preserves SU$(8)$ R-symmetry. Products of the $m_{ijk,a}$ can be chosen to preserve the desired amount of R-symmetry.

All $\mathcal{N}=8$ superamplitudes with $n\geq4$ particles are proportional to the Grassmann delta function $\delta^{(16)}(\widetilde Q)$; the only one that is not is the 3-point anti-MHV superamplitude.

\subsection{3-point EFT Amplitudes}
\label{sec:3pt_amps}

The 3-point amplitudes do not receive corrections in the EFT expansion, since any higher-derivative three-field operator can be recast into higher-point terms via field redefinitions. As a consequence, the 3-point superamplitudes are fixed by little group scaling and are
\be \label{eq:3ptamp}
\mathcal{M}^{\text{MHV}}_3(123)
=\frac{\kappa}{\<12\>^2 \<23\>^2 \<31\>^2}\, 
  \delta^{(16)}\big( \widetilde{Q}\big)\,,
  ~~~~~
  \mathcal{M}^{\text{anti-MHV}}_3
  (123)
  = \frac{\kappa}{[12]^2 [23]^2 [31]^2}\, 
  \prod_{a=1}^8 m_{123,a} \,,
\ee
where $\kappa^2 = 8\pi G_N$, with $G_N$ being Newton's constant. The 3-point superamplitudes are by construction SU$(8)$ invariant. 

\subsection{4-point EFT Amplitudes}
\label{sec:4pt_amps}

At 4 points, we need to consider only one helicity sector, since the anti-MHV amplitudes are equivalent to the MHV ones by conjugation. The MHV superamplitude is
\be \label{eq:4ptamp}
\mathcal{M}^{\text{MHV}}_4(1234)
   = \frac{M_4(--+\,+)}{\<12\>^8} \, 
  \delta^{(16)}\big( \widetilde{Q}\big) \,.
\ee
It is straightforward to show that the most general 4-point EFT amplitude is given by
\be
  M_4(--+\,+) = \<12\>^4 [34]^4 F(s,t,u)\,,
\ee
where $F(s,t,u)$ is fully permutation symmetric in $s$, $t$, and $u$.  
Together with momentum conservation, $s+t+u=0$, this leaves only the monomials $(stu)^{n_1}(s^2+t^2+u^2)^{n_2}$ as independent structures, so 
in the low-energy limit, we can write 
\begin{equation}
\label{eq:Fstu}
    F(s,t,u)=\frac{\kappa^2}{stu} + f(s,t,u)\,,
\end{equation}
with \begin{equation}\label{eq:lowEfparam}
    f(s,t,u)= \sum_{n_1,n_2=0}^\infty g_{n_1,n_2}^{(3n_1 + 2n_2)} \, (stu)^{n_1} (s^2+t^2+u^2)^{n_2}\,.
\end{equation}
The first term in \reef{eq:Fstu} captures the leading-order SUGRA interaction, while $f(s,t,u)$ systematically encodes the contributions of local higher-derivative corrections to the 4-point amplitude. 
The effective couplings $g_{n_1,n_2}^{(k)}$ are in one-to-one correspondence with the Wilson coefficients of the 4-field local operators $D^{2k}R^4$ in the effective action, where we denote $k=3n_1 + 2n_2$. For ease of notation, we  follow~\cite{Caron-Huot:2020cmc,Albert:2024yap,Berman:2024eid} and use $g_{n_1,n_2}^{(k)} \to \{ g_k,g_k',g_k'',\dots\}$ to label the $\lfloor \frac{k+2}{2} \rfloor - \lfloor \frac{k+2}{3} \rfloor$ different Wilson coefficients at each order $s^k$, ordered by increasing $\{n_1,n_2\}$. For example, we relabel $\{ g^{(6)}_{0,3}, g^{(6)}_{2,0}\} \to \{ g_6, g_6'\} $. This way, the first few terms in the low-energy expansion are
\begin{align}
\label{eq:fstu}
    f(s,t,u)=&\, g_{0} + g_{2} \, (s^2+t^2+u^2) + g_{3} \, stu + g_{4} \, (s^2+t^2+u^2)^2 + g_{5} \, stu \,(s^2+t^2+u^2) \nonumber \\[0.1cm] 
    & + g_6 \, (s^2+t^2+u^2)^3 + g_6' \, (stu)^2 + \dots  \,.
\end{align}
As explicit examples of component amplitudes, we have
\begin{equation}
\label{eq: scalar_ampls_SUGRA}
M_4(zz\bar{z}\bar{z}) = s^4 F(s,t,u)\,, \qquad M_4(z\bar{z}z\bar{z}) = t^4 F(s,t,u)\,,
\end{equation}
where we have chosen $z=z^{1234}$ and $\bar{z} = z^{5678}$ as the conjugate scalars. In particular, the amplitude $M_4(zz\bar{z}\bar{z})$, with its explicit overall $s^4$-factor, is well-suited for the S-matrix bootstrap~\cite{Haring:2023zwu,Albert:2024yap}. 

\subsection{5-point EFT Amplitudes}
\label{sec:5pt_amps}

At 5 points, we begin with the MHV superamplitude 
\be
\label{eq:5pt_MHV_superamp}
\mathcal{M}^{\text{MHV}}_5(12345)
   = \frac{M_5(--+++)}{\<12\>^8} \, 
  \delta^{(16)}\big( \widetilde{Q}\big) \,.
\ee
At leading order, $M_5(--+++)$ can be computed with BCFW recursion~\cite{Britto:2004ap,Britto:2005fq}, resulting in
\begin{equation}
\label{eq:LO5ptamp}
M_5^{\text{LO}}(--+++)=-\kappa^3\frac{\ab{12}^8\,\epsilon[1234]}{\ab{12}\ab{13}\ab{14}\ab{15}\ab{23}\ab{24}\ab{25}\ab{34}\ab{35}\ab{45}}\,,
\end{equation}
where $\eps[1234]$ denotes the Levi–Civita contraction
\be
\label{eq:Levi_Civita_def}
 \eps[1234] = 4 i \eps_{\mu \nu \rho \sigma} p_1^\mu p_2^\nu  p_3^\rho  p_4^\sigma =
 [12]\<23\>[34]\<41\>
 -\<12\>[23]\<34\>[41] \,.
\ee 
At higher orders, we can simply write an ansatz for the 5-point amplitude that respects its little group scaling and physical poles. One possibility is the following:
\begin{equation}
\label{eq:5ptansatz}
    M_5(--+++) =   \frac{-\langle 12 \rangle^8 \, }{\langle 12 \rangle \langle 13 \rangle \langle 14 \rangle \langle 15 \rangle 
  \langle 23 \rangle \langle 24  \rangle \langle 25 \rangle
  \langle 34 \rangle \langle 35 \rangle \langle 45 \rangle} \, G_5(12345) \,,
\end{equation}
where $G_5(12345)$ includes both parity-even and -odd terms,
\begin{equation}
\label{eq:def_G5}
    G_5(12345) = V_5(12345) + \epsilon[1234]\,Q_5(12345)\,.
\end{equation}
Here, $V_5(12345)$ and $Q_5(12345)$ are polynomials in the five independent 5-point Mandelstam variables $\{s_{12}, s_{23}, s_{34}, s_{45}, s_{15}\}$. For instance, at the lowest orders, we have
\begin{equation}
\begin{split}
    V_5(12345) =&\, v_{0,1} + v_{1, 1} s_{12} + v_{1, 2} s_{23} + v_{1, 3} s_{34} + v_{1, 4} s_{45} + v_{1, 5} s_{15} + \dots \,, \\[0.2cm]
    Q_5(12345) =&\, q_{0,1} + q_{1, 1} s_{12} + q_{1, 2} s_{23} + q_{1, 3} s_{34} + q_{1, 4} s_{45} + q_{1, 5} s_{15} + \dots \,.
\end{split}
\end{equation}
The parameters $v_{k,r}$ and $q_{k,r}$ denote the $r$ different coefficients at each order $s^k$. Demanding that the 5-point ansatz~\eqref{eq:5ptansatz} has the correct factorization into lower-point amplitudes on the physical poles of $M_5(--+++)$ fixes the 5-point parameters $v_{k,r}$ and $q_{k,r}$ in terms of the 4-point Wilson coefficients and $\kappa^2$, apart from coefficients of local terms; see Appendix~\ref{app:5pt_analysis} for details. The NMHV amplitude can be parameterized similarly.
The resulting amplitudes are compatible with both $\mathcal{N}=8$ SUSY and full SU$(8)$ R-symmetry; indeed, with unbroken R-symmetry, there are no other independent amplitudes at 5 points. 

With $\text{SU}(4) \times\text{SU}(4)$ R-symmetry, there is, in addition to the MHV sector, also the 5-point  N$^{(1,0)}$MHV (along with their conjugates, the NMHV and N$^{(0,1)}$MHV, respectively). The corresponding superamplitude is given by~\cite{Elvang:2010xn}
\begin{equation}
    \mathcal{M}_5^{\text{N}^{(1,0)}\text{MHV}}(12345) = \frac{\delta^{(16)} \big( \widetilde{Q}\big)}{\<45\>^8} \,  Y_{1111} \,
   M_5(z^{1234}++--)\,,
\end{equation}
where 
\begin{equation}
\label{eq:def_Yijkl}
    Y_{ijkl} \equiv \frac{m_{i(n-3)(n-2),1}\,m_{j(n-3)(n-2),2}\,m_{k(n-3)(n-2),3}\,m_{l(n-3)(n-2),4}}{[n-3, n-2]^4} \,,
\end{equation}
with $n=5$ in this case, which depend on the $m_{ijk,a}$, given in~\eqref{eq:mijks}.

In the leading SUGRA theory, the SU$(8)$-violating amplitude $M_5(z^{1234}++--)$ vanishes. Moreover, since there are no SU$(8)$-violating 3- and 4-point amplitudes, the EFT amplitudes in the N$^{(1,0)}$MHV sector cannot have any factorization channels and as a result they must be manifestly local. In the SUSY Ward identity
\begin{equation}
    M_5(z^{1234} - + - +) = \frac{\<24\>^4 [35]^4}{\<45\>^4 [23]^4} \, M_5(z^{1234} ++--)
\end{equation}
both amplitudes have to be local, which is possible only when 
\begin{equation}
    M_5(z^{1234}++--) = \ab{45}^4 \sb{23}^4 \,\widetilde{G}_5(12345)\,,
\end{equation}
for a function $\widetilde{G}_5$ that is a polynomial in the Mandelstam variables. Analogously to~\eqref{eq:def_G5}, it includes both parity-even and -odd terms and these are in one-to-one correspondence with the local 5-point contact interactions in the EFT.

This completes the construction of 5-point superamplitudes in EFTs with $\text{SU}(4) \times\text{SU}(4)$ R-symmetry. We now proceed to 6 points.

\section{6-point NMHV Superamplitude with \texorpdfstring{$\text{SU(4)} \times\text{SU(4)}$}{SU(4)xSU(4)} R-symmetry}
\label{sec:Rsym_SU4xSU4}

In this section, we study the 6-point superamplitudes in the EFT expansion. With $\text{SU}(4) \times\text{SU}(4)$ R-symmetry, the independent helicity sectors at 6-point are MHV, NMHV, N$^{(1,0)}$MHV, and N$^{(2,0)}$MHV, together with their conjugates under $CP$ and $S$ defined in Subsection \ref{ssec:discrete}.    
To set the stage for the nonlinear constraints, which will be derived in the next section by imposing peculiar parity on $(\mathbf{6},\mathbf{6})$ scalar amplitudes, we focus here on NMHV amplitudes and relegate other sectors to Appendix~\ref{app:6ptHelSect}.
For a similar reason, we postpone the discussion of the SU$(8)$ R-symmetry case to the end of this section, where it can be simply obtained by specializing the $\text{SU}(4) \times\text{SU}(4)$ formulas.

Let us briefly comment on the role of the $\mathbb{Z}_2\times \mathbb{Z}_2$ symmetry generated by $CP$ and $S$.  As can be seen from the explicit formulas in the preceding section, 3- and 4-point amplitudes are automatically invariant under both $CP$ and $S$ due to maximal SUSY; see~\eqref{eq:3ptamp}--\eqref{eq:4ptamp}.
This implies that all factorization channels of the 5- and 6-point amplitudes also automatically preserve these symmetries. Hence, the choice of whether or not to impose these symmetries can only affect simple \emph{linear} constraints on the 5-point and 6-point local contact interactions. Since we expect these assumptions to have no significant effect on our main conclusions, we assume $CP$ and $S$ throughout this section for simplicity.

\subsection{6-point NMHV Superamplitude}
\label{sec:SU4xSU4_NMHV_Superamp}

As shown in~\cite{Elvang:2009wd,Elvang:2010kc}, the 6-point NMHV superamplitude with $\text{SU}(4) \times\text{SU}(4)$ R-symmetry can be expressed as
\begin{equation}
\label{CsqNMHV}
    {\mathcal M}_6^\text{NMHV}(123456)=\frac{\delta^{(16)}\bigl(\tilde Q\bigr)}{\<56\>^8 }\!\!\!\!
    \sum_{\substack{1\leq i\leq j\leq k\leq l\leq 2\\[0.1cm] 1\leq i'\leq j'\leq k'\leq l' \leq 2}}\!\!\!\,Y_{(ijkl)}\,\widetilde{Y}_{(i'j'k'l')}\,M_6(\{i\,j\,k\,l\!\mid\!i'\,j'\,k'\,l'\}\, {+}{+}{-}{-})\,,
\end{equation}
where the SU$(4)$ degree-4 Grassmann polynomials $Y_{ijkl}$ are given in~\eqref{eq:def_Yijkl} and the $\tilde Y_{pquv}$ are defined analogously but with dependence on the Grassmann variables from the second SU$(4)$ factor, i.e.~they have $\eta_{ia}$ with $a=5,6,7,8$. The parenthesis in $Y_{(ijkl)}$ is shorthand for the sum over inequivalent permutations of $\{i, j, k, l \}$, while the notation $\{i\,j\,k\,l \!\mid\! i'\,j'\,k'\,l'\}$ indicates the legs on which the eight remaining $\text{SU}(4) \times\text{SU}(4)$ indices $1,2,\ldots,8$ are placed. The $S$ symmetry that interchanges the two SU$(4)$'s implies that
\begin{equation}
  \label{SU4flip}
    M_6(\{i\,j\,k\,l\!\mid\!i'\,j'\,k'\,l'\}\, {+}{+}{-}{-}) 
    = M_6(\{i'\,j'\,k'\,l'\!\mid\!i\,j\,k\,l\}\, {+}{+}{-}{-})\,.
\end{equation}
It then follows from \reef{CsqNMHV} and \reef{SU4flip} that 15 independent amplitudes are needed to specify the 
$\text{SU}(4) \times\text{SU}(4)$-invariant superamplitude, e.g.
\be
 \label{SU4base1}
 \begin{split}
\mathbb{M}_1 &\equiv 
M_6(\{1111|1111\}\, {+}{+}{-}{-})
= M_6({-}{+} {+}{+}{-}{-})\,,
\\
\mathbb{M}_2 &\equiv 
M_6(\{1111|1112\}\, {+}{+}{-}{-})
= M_6(\psi^{1234|567}\psi^{8} {+}{+}{-}{-})\,,
\\
\mathbb{M}_3 &\equiv 
M_6(\{1111|1122\}\, {+}{+}{-}{-})
= M_6(v^{1234|56}v^{78} {+}{+}{-}{-})\,,
\\
\vdots
\\
\mathbb{M}_{6} &\equiv 
M_6(\{1112|1112\}\, {+}{+}{-}{-})
= M_6(v^{123|567}v^{4|8} {+}{+}{-}{-})\,,
\\
\mathbb{M}_{7} &\equiv 
M_6(\{1112|1122\}\, {+}{+}{-}{-})
= M_6(\chi^{123|56}\chi^{4|78} {+}{+}{-}{-})\,,
\\
\vdots
\\
\mathbb{M}_{15} &\equiv 
M_6(\{2222|2222\}\, {+}{+}{-}{-})
= M_6({+}{-} {+}{+}{-}{-})\,.
 \end{split}
\ee

The bottom-up construction of EFT amplitudes is done by explicitly matching residues to products of lower-point amplitudes on each factorization channel and parameterizing  all possible local terms systematically. 
However, the amplitudes in the basis \eqref{SU4base1} are technically challenging to construct from lower-point amplitudes. All we need, though, is a set of 15 independent amplitudes to use as a basis. For technical simplicity we choose to form our basis from
\be
\label{eq:amplitude_S1_basis}
  \mathbb{S}_1 = M_6(z^{1234} z^{1234} z^{1|678} z^{2|578} z^{3|568} z^{4|567})\,.
\ee
The amplitude \reef{eq:amplitude_S1_basis} is invariant under the exchange of the first two identical states, and by relabeling the R-indices it becomes clear that it has full permutation symmetry in the last four external states. It thus enjoys an $S_2 \times S_4$ symmetry. This is why we chose it: the $6!/(2! \, 4!) = 15$ independent permutations $\mathbb{S}_i$ of $\mathbb{S}_1$ can be used to parameterize the 6-particle NMHV amplitude using a single independent function $\mathbb{S}_1$.

Beyond $S_2 \times S_4$ symmetry, $\mathbb{S}_1$ enjoys two other useful properties: it has no 2-particle poles and only involves 3-particle factorization poles, namely in the \mbox{134-,} 135-, 136-, 145-, 146- and 156-channels, and it can be assumed to be parity-symmetric.
The latter follows from $CP$ and the swap $S$ of the two SU$(4)$'s, since the polarizations in \eqref{eq:amplitude_S1_basis} are invariant under $CS$.

To change basis from $\mathbb{M}_i$ to  $\mathbb{S}_i$, we project each of the 15 amplitudes $\mathbb{S}_i$ out from \reef{CsqNMHV}. This results in a linear system
\begin{equation}
\label{eq:changeofbasisSU4xSU4}
\begin{pmatrix} \mathbb{S}_1 \\ \mathbb{S}_2 \\ \vdots \\ \mathbb{S}_{15} \end{pmatrix}
= C_{15 \times 15}
\begin{pmatrix} \mathbb{M}_1 \\ \mathbb{M}_2 \\ \vdots \\ \mathbb{M}_{15} \end{pmatrix},
\end{equation}
with a $15 \times 15$ matrix depending non-trivially on the kinematics. Upon inversion\footnote{In practice, this is done numerically at a given kinematic point.} of $C_{15 \times 15}$, we obtain the $\text{SU}(4) \times\text{SU}(4)$ superamplitude in terms of the basis of 15 scalar amplitudes,
\begin{equation}
\label{M6fromS1toS15}
    {\mathcal M}_6^\text{NMHV}(123456) =  \sum_{i = 1}^{15} C_{i} \,  \mathbb{S}_{i}\,,
\end{equation}
where the $C_{i}$ are spinor-helicity dependent coefficients. We now proceed to construct the EFT amplitude $\mathbb{S}_{1}$ systematically.

\subsection{Construction of the EFT Amplitude \texorpdfstring{$\mathbb{S}_1$}{S1}}
\label{sec:6pt_NMHVeft}

The amplitude $\mathbb{S}_{1}$  defined in \reef{eq:amplitude_S1_basis} has factorizations only in 3-particle channels, and we only need to compute one of the factorization diagrams, since the rest follow from momentum relabelings using the $S_2 \times S_4$ symmetry. Focusing on the 134-channel, we have 
\begin{align}
  \mathbb{S}_1 \Big|_{s_{134} = 0}
  &= \ \begin{tikzpicture}[baseline={([yshift=-0.1cm]current bounding box.center)}] 
	\node[] (a) at (0,0) {};
    \node[] (a1) at ($(a)+(-0.75,0.75)$) {};
    \node[] (a2) at ($(a)+(-1,0)$) {};
    \node[] (a3) at ($(a)+(-0.75,-0.75)$) {};
	\node[] (b) at ($(a)+(3.5,0)$) {};
    \node[] (b1) at ($(b)+(0.75,0.75)$) {};
    \node[] (b2) at ($(b)+(1,0)$) {};
    \node[] (b3) at ($(b)+(0.75,-0.75)$) {};
	\draw[line width=0.2mm] (a.center) -- (a1.center);
    \node[] at ($(a1)+(-0.35,0.2)$) {$1^{1234}$};
    \draw[line width=0.2mm] (a.center) -- (a2.center);
    \node[] at ($(a2)+(-0.45,0)$) {$3^{1678}$};
    \draw[line width=0.2mm] (a.center) -- (a3.center);
    \node[] at ($(a3)+(-0.45,-0.2)$) {$4^{2578}$};
    \draw[line width=0.2mm] (a.center) -- (b.center);
    \draw[line width=0.2mm] (b.center) -- (b1.center);
    \node[] at ($(b1)+(0.45,0.2)$) {$2^{1234}$};
    \draw[line width=0.2mm] (b.center) -- (b2.center);
    \node[] at ($(b2)+(0.45,0)$) {$5^{3568}$};
    \draw[line width=0.2mm] (b.center) -- (b3.center);
    \node[] at ($(b3)+(0.45,-0.2)$) {$6^{4567}$};
    \node[] at ($(a)+(1.1,-0.3)$) {$(-P)^{3456}$};
    \node[] at ($(b)+(-0.9,-0.275)$) {$P^{1278}$};
    \filldraw[gray!25] (a.center) circle (0.35cm);
    \draw[line width=0.2mm] (a.center) circle (0.35cm);
    \filldraw[gray!25] (b.center) circle (0.35cm);
    \draw[line width=0.2mm] (b.center) circle (0.35cm);
    \node[] at ($(a)+(0,0)$) {$M_4$};
    \node[] at ($(b)+(0,0)$) {$M_4$};
\end{tikzpicture} \nonumber \\
  &=
  M_4\big[
  1^{1234} \, 3^{1678} \, 4^{2578} \, (-P)^{3456}
  \big] \,
  \frac{1}{s_{134}} \,
  M_4\big[
  P^{1278} \, 2^{1234} \, 5^{3568} \, 6^{4567}
  \big] \nonumber \\
  &=
  \frac{s_{13}s_{14}s_{25}s_{26}s_{34}^2s_{56}^2}{s_{134}} \,
  F(s_{13},s_{14},s_{34})
  \, 
  F(s_{25},s_{26},s_{56}) \,.
\end{align}
Adding the remaining factorization channels, we can write the ansatz for the 6-point EFT amplitude $\mathbb{S}_1$ as
\be
\label{eq:ansatz_S1}
  \begin{split}
    \mathbb{S}_1 =&\, 
    \bigg(
    \frac{s_{13}s_{14}s_{25}s_{26}s_{34}^2s_{56}^2}{s_{134}} \,
  F(s_{13},s_{14},s_{34})
  \, 
  F(s_{25},s_{26},s_{56})
  + (3 \lra 5)
  + (3 \lra 6)  
  \bigg) \\
  &
  + (1 \lra 2)
  + \text{CT}_6(123456)\,,
  \end{split}
\ee
where $\text{CT}_6(123456)$ parameterizes the 6-point local contact terms. 

Using the low-energy expansion of $F(s,t,u)$ from \reef{eq:Fstu}, we find
\be
 \begin{split}
 &
\frac{s_{13}s_{14}s_{25}s_{26}s_{34}^2s_{56}^2}{s_{134}} \,
  F(s_{13},s_{14},s_{34})
  \, 
  F(s_{25},s_{26},s_{56})
  \\
  &~~~~= \kappa^4
\frac{s_{34}s_{56}}{s_{134}}
  + \kappa^2 \frac{s_{25}s_{26}s_{34}s_{56}^2}{s_{134}}
  f(s_{25},s_{26},s_{56})
    + \kappa^2
    \frac{s_{13}s_{14}s_{34}^2s_{56}}{s_{134}}
  f(s_{13},s_{14},s_{34})\\
  &~~~~~~~~+
\frac{s_{13}s_{14}s_{25}s_{26}s_{34}^2s_{56}^2}{s_{134}} \,
  f(s_{13},s_{14},s_{34})
  \, 
  f(s_{25},s_{26},s_{56})\,.
  \end{split}
\ee 
The terms with $\kappa^4$ add up to be the pure SUGRA leading-order amplitude. The terms at order $\kappa^2$ are linear in the 4-point EFT couplings $g_n, g_n',\dots$, whereas the terms from $f \times f$ provide terms quadratic in $g_n, g_n',\dots$.

We now turn to the contact terms. Just like in the 5-point analysis from Section~\ref{sec:5pt_amps}, a local contact term for a 6-point scalar amplitude generically combines a parity-even and a parity-odd polynomial at every order. The parity-even polynomials depend on the nine 6-point Mandelstam variables, which can be chosen as
\begin{equation}
\label{eq:6pt_Mandelstams}
    \big\{ s_{12},\, s_{23},\, s_{34},\, s_{45},\, s_{56},\, s_{16},\, s_{123},\, s_{234},\, s_{345} \big\}\,.
\end{equation}
By contrast, a 6-point parity-odd term would involve a choice of five independent Levi–Civita contractions
\begin{equation}
\label{eq:6pt_LC}
    \big\{ \epsilon[2345],\, \epsilon[1345],\, \epsilon[1245],\, \epsilon[1235],\, \epsilon[1234] \big\} \,,
\end{equation}
multiplied by a Mandelstam polynomial. 
However, as just mentioned, assuming $CP$ and $S$ there can be no parity-odd contributions to the amplitude $\mathbb{S}_1$.

The contact terms $\text{CT}_6(123456)$ are thus found by enumerating polynomials in the nine Mandelstam variables in \eqref{eq:6pt_Mandelstams} that are invariant under the $S_2 \times S_4$ permutation symmetry of the external momenta. While it is relatively straightforward to generate such polynomials by symmetrizing arbitrary monomials, it is helpful to know beforehand how many independent polynomials actually exist. Using O$(4)$ character technology as explained in Appendix~\ref{app:counting}, we obtained the following generating function which counts the number of independent parity-symmetric
$S_2 \times S_4$ polynomials at each order:
\begin{align} \label{S2 x S4 generating function}
    & \frac{(1 - s^5)(1 - 2s + 3s^2 - s^3 + 4s^4 + 7s^6 - s^7 + 12s^8 + 7s^{10} + 3s^{11} + 2s^{12} + s^{13})}{(1 - s)^3(1 - s^2)(1 - s^3)^2(1 - s^4)^2(1 - s^6)} = \nonumber \\[0.2cm]
    & \quad 1 + s + 4s^2 + 9s^3 + 22s^4 + 43s^5 + 95s^6 + 176s^7 + 335s^8 + 590s^9 + 1024s^{10} + \dots \, .
\end{align}
We have confirmed these numbers up to order $s^8$ by exhausting the list of monomials that can be symmetrized, and counting the independent resulting polynomials using numerical evaluation at random four-dimensional kinematical points.
In practice, it is relatively easy to generate symmetric polynomials and to count linearly independent ones, and we use the analytic result \eqref{S2 x S4 generating function} to know when our basis is complete. We observe that the complete space can be spanned by finding a set of 29 building-block polynomials up to order $s^7$ and taking products of them to obtain the polynomials at order $s^8$ and beyond. The number of independent polynomials is listed in the first row of Tables~\ref{tab:6pt_SU4xSU4SUSY}--\ref{tab:6pt_SU4xSU4}.

\subsection{SUSY and Factorization Constraints}
\label{sec:6pt_SUSY}

We use all 15 permutations of $\mathbb{S}_1$ to construct the ansatz for the 6-point NMHV superamplitude \eqref{M6fromS1toS15}. To test compatibility with SUSY and tree-level factorization, we project out other component amplitudes and impose that 1) they are free of spurious poles and 2) have the correct residues on each physical pole. This turns out to fix many of the free parameters associated with the local contact terms in $\mathbb{S}_1$.
In the following, we focus on the amplitude $M_6(zzz\bar{z}\bar{z}\bar{z})$, where the conjugate scalars are chosen as $z=z^{1234}$ and $\bar{z}=z^{5678}$. While the pole analysis for this particular amplitude only represents a subset of the possible constraints arising from SUSY and tree-level factorization, we find that examining a selection of other amplitudes does not provide any additional constraints. In practice, we employ BCFW shifts~\cite{Britto:2004ap,Britto:2005fq} to facilitate the numerical evaluation of the residues on massless poles while ensuring momentum conservation on the restricted kinematics; see Appendix~A of~\cite{Elvang:2026pmc} for details.

A basic consistency condition arises from the inversion of the $15\times 15$ matrix in the change of basis \reef{eq:changeofbasisSU4xSU4}. When changing basis to $\mathbb{S}_i$, the determinant of the matrix appears in the denominator of the superamplitude, generating potential high-order spurious poles. 
In the case of $M_6(zzz\bar{z}\bar{z}\bar{z})$, its denominator contains a degree-24 irreducible polynomial in the BCFW shift parameter, which must be canceled by the numerator. 
Requiring the residues to vanish to eliminate any spurious poles yields multiple constraints that fix many of the parameters in the $\mathbb{S}_1$ local terms.

Next, we require that the amplitude $M_6(zzz\bar{z}\bar{z}\bar{z})$ has the correct residues on its physical poles. In particular, it has 2- and 3-particle poles in the 14- and 124-channels; the other poles are permutations thereof. We impose that the residue matches the expected product of lower-point amplitudes, and this further constrains the local term parameters in the ansatz for $\mathbb{S}_1$. This is where the 5-point amplitudes enter the analysis. On the 2-particle poles, the residue is the product of a 3-point amplitude and a 5-point amplitude SUSY-related to the MHV amplitude $M_5(--+++)$ constructed in Section~\ref{sec:5pt_amps}.

\begin{table}[t]
\centering
\vspace{4pt}
\begin{tabular}{|l|ccccccccc|}
\hline
\rule{0pt}{0.8\normalbaselineskip}
\!\!\!Order in $\mathbb{S}_1$ & \!$s^0$ & \!$s^1$ & \!$s^2$ & $s^3$ & $s^4$ & $s^5$ & $s^6$ & $s^7$ & $s^8$ \\
\hline
\rule{0pt}{0.8\normalbaselineskip}
\!\!\!Ansatz & \!1 & \!1 & \!4 & 9 & 22 & 43 & 95 & 176 &  335 \\[0.05cm]
\!$\text{SU}(4) \times\text{SU}(4)$ \& pole analysis & \!0 & \!0 & \!0 & 0 & 1 & 0 & 2 & 3 & 10 \\
\hline
\end{tabular}%
\caption{Number of free coefficients in the local terms of the 6-point NMHV amplitude $\mathbb{S}_1$ (i.e.~not counting the 4- and 5-point Wilson coefficients) up to $\mathcal{O}(s^8)$ in the case with $\text{SU}(4) \times\text{SU}(4)$ R-symmetry. The SUSY consistency constraints correspond to the absence of spurious poles and matching the residue on physical factorization poles for component amplitudes, namely $M_6(zzz\bar{z}\bar{z}\bar{z})$.
}
\label{tab:6pt_SU4xSU4SUSY}
\end{table}

We have carried out this pole analysis for $M_6(zzz\bar{z}\bar{z}\bar{z})$ up to order $\mathcal{O}(s^8)$.  Notably, up to and including order $s^3$, the local terms vanish. At order $s^4$, the ansatz for the 6-point local contact terms depends on 22 parameters $w_{4,i}$, e.g.\footnote{Here we use a set of Mandelstam variables different from~\eqref{eq:6pt_Mandelstams} for compactness.}
\begin{equation}
\begin{aligned}
\!\!\!\mathrm{CT}_6(123456)\Big|_{\mathcal{O}(s^4)} \!=\,& \, w_{4,1}\big(s_{34}s_{35}s_{45}s_{345}
+s_{34}s_{36}s_{46}s_{346} +s_{35}s_{36}s_{56}s_{356}
+s_{45}s_{46}s_{56}s_{456}\big)
\\[-2mm]
&~+w_{4,2}\big(s_{36}^2s_{45}^2+s_{35}^2s_{46}^2+s_{34}^2s_{56}^2\big)
\\
&~+w_{4,3}\big(s_{36}^3s_{45}+s_{36}s_{45}^3+s_{35}^3s_{46}+s_{35}s_{46}^3+s_{34}^3s_{56}+s_{34}s_{56}^3\big)
\\
&~+w_{4,4}\big(s_{35}s_{36}s_{45}s_{46}+s_{34}s_{36}s_{45}s_{56}+s_{34}s_{35}s_{46}s_{56}\big)
+\dots \,.
\end{aligned}
\end{equation}
Requiring that the amplitude has the correct residues reduces the number of free parameters to just one; all of the other coefficients of the contact terms are set to linear combinations of $g_0$ and $w_{4, 1}$:
\begin{equation}
\begin{aligned}
\label{eq:solution_factorization_SU4xSU4}
    w_{4, 2} &= 2 \kappa^2 g_0 + \frac{8}{3} w_{4, 1}\,, \qquad \ 
    w_{4, 3} = -\kappa^2 g_0 - \frac{2}{3} w_{4, 1}\,, \qquad \ \hspace{0.05cm}
    w_{4, 4} = -8 \kappa^2 g_0 - \frac{8}{3} w_{4, 1}\,,\\
    w_{4, 5} &= -\kappa^2 g_0 + \frac{2}{3} w_{4, 1}\,, \qquad
    w_{4, 6} = -3 \kappa^2 g_0 - \frac{1}{3} w_{4, 1}\,, \qquad w_{4, 7} = \frac{4}{3} w_{4, 1}\,,\\
    w_{4, 8} & = \ldots = w_{4, 22} = 0\,.
\end{aligned}
\end{equation}
Next, at order $s^5$, there are no pole terms in $\mathbb{S}_1$, and  all coefficients of the local terms vanish by the pole analysis. At orders $s^6$ and beyond, only a few free parameters remain in the local terms of $\mathbb{S}_1$; the rest are determined by the 4-point EFT couplings $g_n$, $g_n'$, \dots, and $\kappa$. The number of free parameters in the $\mathbb{S}_1$ local terms is summarized in Table~\ref{tab:6pt_SU4xSU4SUSY}.

\subsection{Imposing SU(8) R-symmetry}
\label{sec:wSU(8)}

Let us briefly consider the case with full SU$(8)$ R-symmetry. The construction of the ansatz for the $\mathbb{S}_1$ amplitude is not sensitive to the difference between SU$(8)$ and SU$(4)\times$SU$(4)$. 
However, with unbroken SU$(8)$ symmetry, the scalars are all on equal footing. 
For example, it requires equality of the NMHV amplitudes 
\be
 M_6(z^{1234} z^{5678}--++)
 =
 - M_6(z^{123|5} z^{4|678}--++)
 =
 M_6(z^{12|56} z^{34|78}--++)\,,
\ee
where the relative minus sign comes from the Grassmann derivatives needed to project out the component amplitudes from the superamplitude. Imposing equivalence of the scalar amplitudes further constrains the remaining free parameters of the local terms in $\mathbb{S}_1$. 

\begin{table}[t]
\centering
\vspace{4pt}
\resizebox{\textwidth}{!}{\begin{tabular}{|l|ccccccccccccc|}
\hline
\rule{0pt}{0.8\normalbaselineskip}
\!\!Order in $\mathbb{S}_1$ & $s^0$ & $s^1$ & $s^2$ & $s^3$ & $s^4$ & $s^5$ & $s^6$ & $s^7$ & $s^8$  & $s^9$ & $s^{10}$ & $s^{11}$ & $s^{12}$ \\
\hline
\rule{0pt}{0.8\normalbaselineskip}
\!\!Ansatz & 1 & 1 & 4 & 9 & 22 & 43 & 95 & 176 &  335 & 590 & 1024 & 1686 & 2743 \\[0.05cm]
$\text{SU}(8)$ \& pole analysis & 0 & 0 & 0 & 0 & 0 & 0 & 0 & 0 & 7 & 18 & 52 & 112 & 242 \\
\hline
\end{tabular}}%
\caption{Number of free coefficients (other than the 4- or 5-point Wilson coefficients) in the ansatz of the 6-point NMHV amplitude $\mathbb{S}_1$ up to $\mathcal{O}(s^{12})$, and the reduction after imposing SU$(8)$ R-symmetry and the corresponding SUSY Ward identity.}
\label{tab:6pt_SU8}
\end{table}

Alternatively, one can directly impose SU$(8)$ R-symmetry on the superamplitude construction; see Appendix~\ref{app:Rsym_SU8} for details. Since the SU$(8)$-superamplitude is fully determined by just nine basis amplitudes, we can use a subset of nine amplitudes from the fifteen permutations of $\mathbb{S}_1$. Projecting out the remaining six $\mathbb{S}_i$'s then results in nontrivial consistency conditions for the parameterization of $\mathbb{S}_1$. These consistency conditions correspond to different instances of SUSY Ward identities~\cite{Grisaru:1976vm,Grisaru:1977px}. For example, we have
\begin{equation}
\label{eq:equatingpermsSU8}
    \mathbb{S}_{10} = \sum_{i}^{9} c_{i} \, \mathbb{S}_{i}\,,
\end{equation}
where the $c_i$ depend only on spinor-helicity variables. As summarized in Table~\ref{tab:6pt_SU8}, solving these identities completely fixes $\mathbb{S}_1$ in terms of the 4-point Wilson coefficients, with no free parameters in the local terms, up to and including order $s^7$. For example, at order $s^4$ the remaining free parameter from~\eqref{eq:solution_factorization_SU4xSU4} is fixed to be $w_{4,1}= - \frac{6}{5} \kappa^2 g_{0}$. Starting at order $s^8$, we still have free parameters in the parameterization of the local contact terms. Still, the SU$(8)$ amplitudes are significantly more constrained than when the R-symmetry is relaxed to $\text{SU}(4) \times\text{SU}(4)$.

Lastly, imposing either peculiar parity or vanishing single soft scalar limits to order $s^7$ requires $g_0=g_2=g_3 =0$, and at higher orders we similarly find that the 4-point Wilson coefficients are forced to vanish, even in the presence of free 6-point coefficients in the amplitude. Indeed, positivity (see Section \ref{sec:positivity}) together with $g_0 = 0$ implies that all of the 4-point Wilson coefficients must vanish. Hence, there is no unitary SU$(8)$-invariant $\mathcal{N}=8$ SUSY completion of the tree-level, leading-order SUGRA theory for which the vanishing single soft scalar limits or the peculiar parity property is preserved.

\section{Consequences of Peculiar Parity}
\label{sec:PP}

In this section, we study the consequences of imposing peculiar parity on the 6-point amplitudes with external scalars from the (\textbf{6},\textbf{6}) irrep. 
We also discuss single-soft scalar limits. Finally, we discuss examples of amplitudes that are compatible or incompatible with the nonlinear constraints required by peculiar parity.

\subsection{Parity and Nonlinear Constraints}
\label{sec:Nonlinear_constraints}

Consider the six-scalar amplitude 
\begin{equation}
\label{eq:def_Z1,1}
  \mathbb{Z}_{1,1} \equiv M_6(z^{12|56} z^{13|57} z^{14|58} z^{23|67} z^{34|78} z^{24|68})\,
\end{equation}
with all scalars from the (\textbf{6},\textbf{6}) scalar sector. This amplitude can be viewed as the double copy of two  $\mathbb{Z}_1$ amplitudes, where  $\mathbb{Z}_1$ is the six-scalar amplitude studied in the $\mathcal{N}=4$ SYM EFT analysis of \cite{Elvang:2026pmc}. We do {\em not} construct 
$\mathbb{Z}_{1,1}$ as a double-copy here; rather we project it out from the 6-point superamplitude~\eqref{M6fromS1toS15}, given by the basis of the fifteen $\mathbb{S}_i$ amplitudes. 

For the $\mathbb{Z}_{1,1}$ amplitude, the peculiar parity condition is the requirement that 
\begin{equation}
\label{eq:parity_condition_Z1,1}
    \mathbb{Z}_{1,1} = \overline{\mathbb{Z}}_{1,1}\,,
\end{equation}
where $\overline{\mathbb{Z}}_{1,1} \equiv M_6(z^{34|78} z^{24|68} z^{23|67} z^{14|58} z^{12|56} z^{13|57})$ is the $C$-conjugate of $\mathbb{Z}_{1,1}$. Since we are assuming $CP$, this is equivalent to imposing spacetime parity $P$ on $\mathbb{Z}_{1,1}$.
That is, we require symmetry in the exchange of angle and square brackets, i.e.~the absence of parity-odd Levi–Civita contractions.\footnote{One may of course also ask what happens if peculiar parity is imposed on six-scalar amplitudes involving the axio-dilaton or the $(\mathbf{4},\mathbf{\bar{4}})$ scalars (and their conjugates). 
In those cases, we find that peculiar parity requires $g_0=0$ at order $\mathcal{O}(s^4)$ in the EFT expansion. When combined with positivity (see Section \ref{sec:positivity}), the condition $g_0 = 0$ implies the absence of a UV completion, and only allows the leading-order SUGRA theory. Hence, we focus on peculiar parity only for 6-point amplitudes with external scalars from the $(\mathbf{6},\mathbf{6})$ sector.}

The parity condition~\eqref{eq:parity_condition_Z1,1} cannot be satisfied for arbitrary values of the 4-point Wilson coefficients, but requires nonlinear constraints among them. Imposing  \eqref{eq:parity_condition_Z1,1} order by order in the EFT expansion up to $\mathcal{O}(s^{13})$, we find the following nonlinear relations:
{\allowdisplaybreaks
\begin{align}
\mathcal{O}(s^7)\!: \quad & \kappa^2 g_3 = \frac{1}{2}g_0^2\,, \nonumber \\[1.5mm]
\mathcal{O}(s^9):\! \quad & \kappa^2 g_5 = g_2 g_0 \,, \nonumber \\[1.5mm]
\mathcal{O}(s^{10})\!: \quad &\kappa^4g_6' = \frac{8}{3}\kappa^4 g_6 + \frac{1}{6}g_0^3\,, \nonumber \\[1.5mm]
\mathcal{O}(s^{11})\!: \quad & \kappa^2g_7 = g_4g_0 + \frac{1}{2}g_2^2 \,, \nonumber \\[1.5mm]
\mathcal{O}(s^{12})\!: \quad & \kappa^4 g_8' = 8 \kappa^4 g_8 + \frac{1}{2} g_2 g_0^2 \,, \nonumber \\[1.5mm]
\mathcal{O}(s^{13})\!: \quad & \kappa^2 g_9 = g_6 g_0 + g_4 g_2 \,, \qquad \kappa^6 g_9' = \frac{8}{3} \kappa^4 g_6 g_0 + \frac{1}{24} g_0^4 \,. \nonumber \\[-0.3cm]
\label{eq:nonlinear_constraints_gk}
\end{align}
}
These relations are derived bottom-up, and must therefore be satisfied by any UV completion to SUGRA consistent with $\mathcal{N}=8$ SUSY, tree-level factorization, and the peculiar parity condition~\eqref{eq:parity_condition_Z1,1}.

The same conditions also fix most of the free coefficients from the $\mathbb{S}_1$ ansatz~\eqref{eq:ansatz_S1} in terms of the 4-point Wilson coefficients. In practice, it is a lot simpler to impose the peculiar parity condition than to check the pole structure, so we can go to higher orders. The second row in Table~\ref{tab:6pt_SU4xSU4} lists the number of free variables after only imposing \eqref{eq:parity_condition_Z1,1}, without the pole analysis from Section \ref{sec:6pt_SUSY}. The amplitudes resulting from the parity constraints are fully fixed in terms of $g_n$, $g_n'$,\dots, and $\kappa$ up to and including order $s^7$, and it is easy to check to that order that the amplitudes are free of spurious poles and have the correct residues on the physical poles. Starting at order $s^8$, however, carrying out the pole analysis as in Section~\ref{sec:6pt_SUSY} further reduces the number of free coefficients in the 6-point local terms of $\mathbb{S}_1$; for example at order $s^8$ there are only two coefficients left free, not five. The number of free parameters in the ansatz of the local terms of $\mathbb{S}_1$ after the pole analysis is given in the third row of Table \ref{tab:6pt_SU4xSU4}.

\begin{table}[t]
\centering
\vspace{4pt}
\resizebox{\textwidth}{!}{\begin{tabular}{|l|cccccccccccccc|}
\hline
\rule{0pt}{0.8\normalbaselineskip}
\!\!\!Order in $\mathbb{S}_1$ & \!$s^0$ & \!$s^1$ & \!$s^2$ & $s^3$ & $s^4$ & $s^5$ & $s^6$ & $s^7$ & $s^8$  & $s^9$ & $s^{10}$ & $s^{11}$ & $s^{12}$ & $s^{13}$ \\
\hline
\rule{0pt}{0.8\normalbaselineskip}
\!\!\!Ansatz & \!1 & \!1 & \!4 & 9 & 22 & 43 & 95 & 176 &  335 & 590 & 1024 & 1686 & 2743 & 4276 \\[0.05cm]
\!Peculiar parity & \!0 & \!0 & \!0 & 0 & 0 & 0 & 0 & 0 & 5 & 8 & 28 & 64 & 143 & 274 \\[0.05cm]
\!Pole analysis & \!0 & \!0 & \!0 & 0 & 0 & 0 & 0 & 0 & 2 & 0 & 4 & 6 & 17 & 32 \\
\hline
\end{tabular}}%
\caption{Number of free coefficients (other than the 4- or 5-point Wilson coefficients) in the ansatz of the 6-point NMHV amplitude $\mathbb{S}_1$ up to $\mathcal{O}(s^{13})$ in the case with $\text{SU}(4) \times\text{SU}(4)$ R-symmetry. The count of such free parameters in the first row is for the general ansatz, the second row after imposing the peculiar parity condition~\eqref{eq:parity_condition_Z1,1}, which also leads to the nonlinear constraints among the 4-point Wilson coefficients $g_{k}$, and in the last row after the pole analysis from Section~\ref{sec:6pt_SUSY}.
}
\label{tab:6pt_SU4xSU4}
\end{table}

We note that the condition~\eqref{eq:parity_condition_Z1,1} goes beyond imposing peculiar parity for this scalar amplitude and extends to other scalar amplitudes with 6 external scalars of the type $z^{ab|cd}$. In fact, it even has a further extension that is most easily formulated in terms of the double copy. In that language, the $\mathcal{N}=8$ scalars $z^{ab|cd}$ can be thought of as the double-copies of two $\mathcal{N}=4$ scalars: $z^{ab|cd} = s^{ab} \otimes s^{cd}$. Conjugation within each $\mathcal{N}=4$ factor takes, e.g., $s^{12} \lra s^{34}$ or $s^{56} \lra s^{78}$. Then, we find that peculiar parity extends to each  $\mathcal{N}=4$ factor. 
For instance, a scalar amplitude such as $M_6(z^{12|56} z^{12|57} z^{34|78} \ldots )$ is equal to $M_6(z^{34|56} z^{34|57} z^{12|78} \ldots )$, where we conjugate the $\mathcal{N}=4$ scalar $s^{12} \leftrightarrow s^{34}$ in all instances where it appears. Consequently, imposing the parity condition~\eqref{eq:parity_condition_Z1,1} enhances the SU$(4) \times$SU$(4) \sim$ SO$(6) \times $SO$(6)$ R-symmetry to O$(6) \times$O$(6)$. This is clearly reminiscent of the analysis in $\mathcal{N}=4$ SUSY from~\cite{Elvang:2026pmc}, where an analogous parity condition enhances the SU$(4)$ $\sim$ SO$(6)$ R-symmetry to O$(6)$, which allows for parity transformations in the individual scalars. In this sense, the condition $\overline{\mathbb{Z}}_{1,1} = \mathbb{Z}_{1,1}$ actually can be thought of as
\be
  \mathbb{Z}_{\overline{1},\overline{1}}
 = \mathbb{Z}_{1,1}\,,
\ee
where the bars on the 1's refer to conjugation in the two respective  $\mathcal{N}=4$ states, and that it extends to
\be
  \mathbb{Z}_{1,\overline{1}}
 = \mathbb{Z}_{1,1}
 = \mathbb{Z}_{\overline{1},1}\,.
\ee
We emphasize that the double copy was \emph{not} assumed in our analysis; it is only a convenient tool for labeling the external states and formulating the conjugation properties.

\subsection{Soft Limits}
\label{sec:Soft_limits}

Let us now consider the properties of the EFT amplitudes when a single scalar is taken soft. 
Consider first the amplitudes resulting from imposing peculiar parity: they are completely fixed up to $s^7$ and we find that they automatically satisfy vanishing soft limits for the complex scalars of the type $z^{ab|cd}$, e.g.
\begin{align}
\label{eq:GoldstoneAmpSoftness}
    \underset{p_1\to0}{\lim}\;M_6(z^{12|56}z^{34|78}++--)=0\,.
\end{align}
In fact, even with the unfixed constants in the local terms of $\mathbb{S}_1$ starting at order $s^8$, the scalar soft limit~\reef{eq:GoldstoneAmpSoftness} is automatically satisfied, as checked up to order $s^{13}$. 
However, the vanishing soft scalar limits do not extend to the other two types of scalars. 

We can also reverse the question and ask what constraints soft limits impose on the more general $\text{SU}(4) \times \text{SU}(4)$ amplitudes before requiring peculiar parity. The starting point is the amplitudes obtained from the pole analysis in Section \ref{sec:6pt_SUSY}. Working bottom-up, if we demand all three sets of scalars --- $z^{1234}$, $z^{abc|d}$, and  $z^{ab|cd}$ --- have vanishing single soft limits, or even just two of these types, then at order $s^4$ we must have $g_0=0$. By positivity, this sets all higher-derivative corrections to zero. 

Similarly, we can study the constraints imposed by the soft theorem \eqref{eq:GoldstoneAmpSoftness} \emph{without} any parity assumptions. At $\mathcal{O}(s^4)$, we find that the soft theorem fixes the one free 6-point contact term, so that the 6-point amplitude is completely fixed. However, at $\mathcal{O}(s^7)$ the soft theorem does not fix the contact terms of 
$\mathbb{S}_1$ entirely and, in particular, does not yield the nonlinear constraint between $g_3$ and $g_0$ from~\eqref{eq:nonlinear_constraints_gk}. 

To summarize, peculiar parity ensures the vanishing soft theorems for the scalars of the type $z^{ab|cd}$, but the converse is not true: a vanishing soft limit does not imply the scalar parity condition nor the nonlinear constraints. It is nevertheless possible that imposing the non-vanishing dilaton soft theorems~\cite{DiVecchia:2015oba,DiVecchia:2017gfi} could further restrict the amplitude.

\subsection{Amplitude Examples}
\label{sec:consequences}

In this section, we consider different examples of UV completions that yield well-behaved and SUSY compatible 4-point amplitudes of the form
\be
  M_4(--+\,+) = \<12\>^4 [34]^4 F(s,t,u)\,,
\ee
and test whether they satisfy the nonlinear constraints obtained via higher-point factorization and peculiar parity.

\subsubsection{Closed Superstring Tree-Level Amplitude}
\label{sec:closed_string}

The 4-point closed superstring graviton amplitude is the Virasoro–Shapiro amplitude~\cite{Virasoro:1969me,Shapiro:1970gy} 
\begin{equation}
\label{eq:Virasoro-Shapiro}
   M_4^{\text{str}}(--+\,+) =
  \langle12 \rangle^4 [34]^4\,\frac{\kappa^2}{stu}
\,\frac{\Gamma(1{-}\alpha's) \Gamma(1{-}\alpha't)\Gamma(1{-}\alpha' u)}{\Gamma(1{+}\alpha's) \Gamma(1{+}\alpha't)\Gamma(1{+}\alpha' u)} \,.
\end{equation}
Expanding at low-energy $s\alpha'$, $t\alpha'$, $u\alpha' \ll 1$, the lowest-order Wilson coefficients are
\begin{equation}
\label{eq:Wilson_coeffs_VS}
\begin{split}
    g_0 =&\, 2 \kappa^2 \alpha'^3 \zeta_3\,, \hspace{2.82cm} g_2=\kappa^2 \alpha'^5 \zeta_5\,, \hspace{1.34cm} g_3 = 2\kappa^2 \alpha'^6 \zeta_3^2\,, \qquad\qquad g_4 = \frac{1}{2} \kappa^2 \alpha'^7 \zeta_7\,, \\[1.5mm]
    g_5 =&\,2 \kappa^2 \alpha'^8 \zeta_3 \zeta_5\,, \hspace{2.47cm} g_6 = \frac{1}{4} \kappa^2  \alpha'^9 \zeta_9\,, \hspace{1.08cm} g_6' = \kappa^2 \alpha'^9 \left( \frac{4}{3} \zeta_3^3 + \frac{2}{3} \zeta_9 \right), \\[1.5mm]
    g_7 =&\, \kappa^2 \alpha'^{10} \left( \frac{1}{2} \zeta_5^2 + \zeta_3 \zeta_7 \right), \qquad g_8 = \frac{1}{8} \kappa^2 \alpha'^{11} \zeta_{11}\,, \qquad g_8' = \kappa^2 \alpha'^{11} \left( 2 \zeta_3^2 \zeta_5 + \zeta_{11} \right), \ \ \dots,
\end{split}
\end{equation}
where the Riemann zeta values are $\zeta_b = \sum_{n=1}^\infty 1/n^b$.
The Wilson coefficients $g_k$, $g_k'$,~\dots, have uniform transcendental weight $k+3$.

It is straightforward to check that the Wilson coefficients \reef{eq:Wilson_coeffs_VS} satisfy the nonlinear constraints~\eqref{eq:nonlinear_constraints_gk}. Furthermore, the Wilson coefficients $g_{2m}$ that remain {\em unfixed} by the nonlinear constraints precisely correspond to the first appearance of each $\zeta_\text{odd}$, specifically 
\begin{equation}
\label{eq:g2m_VS}
    g_{2m} = \frac{1}{2^{m-1}} \kappa^2 \alpha'^{2m+3} \zeta_{2m+3}\,, \qquad \text{for} \quad m=0,1,2,\dots\,.
\end{equation}

Our analysis fixes the 5- and 6-point EFT amplitudes at the lowest orders in terms of the 4-point Wilson coefficients and $\kappa^2$. Comparing to the closed string 5- and 6-point amplitudes, we have checked  agreement up to and including $\mathcal{O}(s^7)$ and $\mathcal{O}(s^8)$, respectively, see Appendix~\ref{app:comparison_string} for details. At order $\mathcal{O}(s^{12})$, the 5- and 6-point closed superstring amplitudes feature the first multi-zeta value $\zeta_{3,3,5}$, which is algebraically independent from the single-zeta values. Therefore, starting at that order we expect to have free parameters that are not possible to fix algebraically, even by studying SUSY and factorization of amplitudes with more than six external states.

Let us finally note that the Virasoro–Shapiro amplitude can also be written in an exponentiated form~\cite{Schlotterer:2012ny}:
\begin{equation}
\label{eq:Virasoro-Shapiro-exp}
   M_4^{\text{str}}(--+\,+) =
  \langle12 \rangle^4 [34]^4\,\frac{\kappa^2}{stu}
  \exp\bigg(
  2\sum_{k=0}^\infty
  \frac{\zeta_{2k+3}}{2k+3}{\alpha'}^{2k+3} 
  \big( 
    s^{2k+3}
    +t^{2k+3}
    +u^{2k+3}
  \big)
  \bigg)
 \,.
\end{equation}
This makes the dependence on the odd zeta values manifest and it will be useful for us later. 

\subsubsection{Infinite Spin Tower}
\label{sec:IST}

Another example is the so-called Infinite Spin Tower (IST), which describes the exchange of states with arbitrary spin $J=0,1,2,\dots,$ and the same mass $m$. This type of amplitude frequently appears in the tree-level S-matrix bootstrap, see e.g.~\cite{Caron-Huot:2020cmc,Albert:2022oes,Berman:2023jys,Berman:2024eid,Berman:2025owb,Elvang:2026pmc}, and in the $stu$-symmetric case it is given by
\begin{equation}
\label{eq:M4IST}
     M_4^\text{IST}(--+\,+) = 
\<12\>^4 [34]^4
  \bigg(    
  \frac{\kappa^2}{stu}
  +
\frac{\lambda^{2}}
{(m^2-s)(m^2-t)(m^2-u)}
 \bigg) \,,
\end{equation}
where $\lambda$ sets the scale of the 3-point coupling constant. Expanding at low energies, the first Wilson coefficients are 
\begin{equation}
\begin{split}
    g_0 =&\, \frac{\lambda^2}{m^6}\,, \qquad \qquad g_2=\frac{\lambda^2}{2m^{10}}\,, \qquad \qquad g_3=\frac{\lambda^2}{m^{12}}\,, \qquad \qquad g_4=\frac{\lambda^2}{4m^{14}}\,, \\
    g_5 =&\, \frac{\lambda^2}{m^{16}}\,, \qquad\hspace{-0.15cm} \qquad g_6=\frac{\lambda^2}{8m^{18}}\,, \qquad \qquad g_6'=\frac{\lambda^2}{m^{18}}\,, \qquad \qquad g_7=\frac{3\lambda^2}{4m^{20}}\,.
\end{split}
\end{equation}
When $\lambda^2= 2\kappa^2$ the IST amplitude satisfies all nonlinear constraints. To understand why this is the case, note that for $\lambda^2=2\kappa^2$, the amplitude~\eqref{eq:M4IST} can be rewritten as
\begin{equation}
\label{eq:MIST_rewritten}
     M_4^\text{IST}(--+\,+) =   \langle12 \rangle^4 [34]^4\,\frac{\kappa^2}{stu}
\, \frac{(m^2+s)(m^2+t)(m^2+u)}{(m^2-s)(m^2-t)(m^2-u)} \,.
\end{equation}
This expression is related to the Virasoro–Shapiro amplitude~\eqref{eq:Virasoro-Shapiro} in the following way. 
Setting $\alpha' = 1/m^2 = 1$ for simplicity and using
\begin{equation} \label{zeta 1}
    \frac{\Gamma(1{-}z)}{\Gamma(1{+}z)} = \exp\left(2\gamma \, z + 2 \sum_{k=1}^\infty \frac{\zeta_{2k+1}}{2k+1} \,z^{2k+1}\right)
    ~~\xrightarrow{\zeta_j \to 1} ~~
\frac{1+z}{1-z}\,e^{2(\gamma-1)z}
\end{equation}
for each of the three $\Gamma$-function ratios in Virasoro–Shapiro,  
we find the expression for the IST amplitude \reef{eq:MIST_rewritten}; the Euler-Mascheroni constant $\gamma$ drops out because $s+t+u=0$. 
Thus, the Virasoro–Shapiro amplitude with all $\zeta_{k}$ formally replaced by 1 gives the IST amplitude! This explains why the IST satisfies all nonlinear constraints.

Note that \eqref{zeta 1} makes sense because only odd zeta values appear in the ratio of $\Gamma$-functions, and these are believed to be algebraically independent. The substitution would not make sense for an individual $\Gamma$-function since the exponent then contains even zeta values, which satisfy algebraic relations incompatible with $\zeta_{2k}\to 1$.  Thus it is only for the closed string that one can formally obtain the IST from the string amplitude.
This is consistent with the $\mathcal{N}=4$ SYM analysis~\cite{Elvang:2026pmc}, where it was found that nonlinear constraints rule out the $s \lra u$ symmetric IST, essentially because the corresponding nonlinear constraints are tuned to the non-trivial relations among the even zeta values that appear in the Veneziano amplitude.

It is unclear whether \eqref{eq:MIST_rewritten}
describes only a mathematical model of a 4-point amplitude, or if it is part of a more complete theory.
An analogous non-gravitating IST amplitude was recently argued to be the limit of a unitary UV complete theory~\cite{Calisto:2026cdy}.

\subsubsection{Generalized Spin Towers}
The relationship between the IST and the Virasoro–Shapiro amplitudes can also be viewed from a different perspective, which allows us to determine an infinite family of amplitudes that also satisfy both positivity and the nonlinear constraints. This comes from the infinite product form of the Virasoro–Shapiro amplitude \cite{Virasoro:1969me,Shapiro:1970gy},
\begin{align}
\label{eq:product_form_VS}
M_4^{\text{str}}(--+\,+)=  \langle12 \rangle^4 [34]^4\,\frac{\kappa^2}{stu}\prod_{n=1}^{\infty}\frac{(n+\alpha's)(n+\alpha't)(n+\alpha'u)}{(n-\alpha's)(n-\alpha't)(n-\alpha'u)}\,.
\end{align}
Consider the amplitude defined by the finite truncation of this infinite product and with $\alpha'=1/m^2$:
\begin{align}
M_4^{(N)}(--+\,+)=  \langle12 \rangle^4 [34]^4\,\frac{\kappa^2}{stu}\prod_{n=1}^{N}\frac{(nm^2+s)(nm^2+t)(nm^2+u)}{(nm^2-s)(nm^2-t)(nm^2-u)}\,.
\end{align}
With $N = 1$, we get the product form of the IST amplitude \eqref{eq:MIST_rewritten}. For positive integer $N$, the spectrum of the amplitude is clear: there is an infinite tower of states at $n m^2$ for integers $1 \leq n \leq N$. 

This amplitude can also be extended to non-integer $N$, as the product combines into a ratio of $\Gamma$-functions multiplying the string amplitude with $\alpha'=1/m^2$:
\begin{align}\label{eq:M4N}
    M_4^{(N)}(--+\,+)=M_4^{\text{str}}(--+\,+)\,\frac{\Gamma(1+s/m^2+N) \Gamma(1+t/m^2+N)\Gamma(1+u/m^2+N)}{\Gamma(1-s/m^2+N) \Gamma(1-t/m^2+N)\Gamma(1-u/m^2+N)} \,.
\end{align}
This form of the amplitude makes it clear that there is a simple analytic continuation in $N$ for arbitrary real $N>1$. (For complex $N$ or $N<1$, the function has poles at complex values of $s$, which is non-physical, or values of $s < m^2$, which violates our assumption that $m^2$ is the lightest (nonzero) mass at which states are exchanged.)  Equation~\eqref{eq:M4N} also manifests the spectrum of the amplitude for non-integer $N$: there are spin towers at $s = nm^2$ for any positive integer $n$ and also at $(n+N)m^2$ for $N\in \mathbb{R}^{+}\setminus\mathbb{Z}$. 

The corresponding low-energy coefficient can be expressed in terms of generalized harmonic numbers,
\begin{equation}
\label{eq:Wilson_coeffs_MN}
\begin{split}
    g_0 =&\, 2 \frac{\kappa^2}{m^6} H_{N}^{(3)}\,, \hspace{1.32cm} g_2=\frac{\kappa^2}{m^{10}} H_{N}^{(5)}\,, \hspace{0.65cm} g_3 = 2\frac{\kappa^2}{m^{12}} (H_{N}^{(3)})^2\,, \qquad g_4 = \frac{1}{2} \frac{\kappa^2}{m^{14}} H_{N}^{(7)}\,, \\[1.5mm]
    g_5 =&\,2 \frac{\kappa^2}{m^{16}} H_{N}^{(3)} H_{N}^{(5)}\,, \hspace{0.4cm} g_6 = \frac{1}{4} \frac{\kappa^2}{m^{18}}  H_{N}^{(9)}\,, \hspace{0.4cm} g_6' = \frac{\kappa^2}{m^{18}} \left( \frac{4}{3} (H_{N}^{(3)})^3 + \frac{2}{3} H_{N}^{(9)} \right), \ \dots \ ,
\end{split}
\end{equation}
i.e.~one simply takes the zeta values $\zeta_k$ in the string amplitude~\eqref{eq:Wilson_coeffs_VS} and replaces them with $H_{N}^{(k)}$. The generalized harmonic numbers are defined by
\begin{align}
    H_N^{(k)} \equiv \sum_{n =1}^{N}\frac{1}{n^{k}}\,,
\end{align}
with $\lim_{N\to\infty}H_N^{(k)} = \zeta_{k}$ as expected. 

Since this replaces the odd zeta values with the harmonic numbers, the coefficients of $M_4^{(N)}$ clearly satisfy all of the nonlinear constraints~\eqref{eq:nonlinear_constraints_gk}. In fact, it turns out that any function with a product form like \reef{eq:M4N} for any (finite) set of masses, whether sequential integers or not, also obeys the nonlinear constraints. However, not all such amplitudes have positive partial wave expansions.  As will be discussed in the next section, the $M_4^{(N)}$ appear to provide a continuous family of positive amplitudes
that interpolate between the IST and Virasoro–Shapiro amplitudes. Among other product amplitudes, the latter is singled out by having states of only finite spin at finite mass levels.

\subsubsection{Exchange of a Massive \texorpdfstring{$\mathcal{N}=8$}{N=8} Supermultiplet}
\label{sec:massive_exchange}

Lastly, consider the following 4-point amplitude,
\begin{equation}
     M_4(--+\,+) = 
\<12\>^4 [34]^4
  \bigg(    
  \frac{\kappa^2}{stu}
  +
\frac{\lambda^{2}}
{m^2-s} +
\frac{\lambda^{2}}
{m^2-t} +
\frac{\lambda^{2}}
{m^2-u}
 \bigg) \,.
\end{equation}
This describes the exchange of a massive $\mathcal{N}=8$ supermultiplet with mass $m$, coupled to the massless states with coupling constant $\lambda$. In this particular amplitude, a spin-0 state of the massive supermultiplet is exchanged in the $s$-channel, while a spin-4 state is exchanged in the $t$- and $u$-channels. The first few Wilson coefficients are
\begin{equation}
    g_0 = \frac{3\lambda^2}{m^2}\,, \qquad g_2 = \frac{\lambda^2}{m^6}\,, \qquad g_3 = \frac{3\lambda^2}{m^8}\,, \qquad g_4 = \frac{\lambda^2}{2m^{10}}\,, \qquad g_5 = \frac{5\lambda^2}{2m^{12}}\,, \ \dots \ .
\end{equation}
In this case, the first two nonlinear constraints from~\eqref{eq:nonlinear_constraints_gk} cannot simultaneously be  satisfied for any nonzero choice of the couplings $\lambda$ and $\kappa$. Thus, this theory is not consistent with the joint constraints of $\mathcal{N}=8$ SUSY, tree-level factorization at 6 points and the peculiar parity condition.

\section{Numerical and Analytic Bootstrap}
\label{sec:positivity}

We now incorporate the nonlinear constraints \eqref{eq:nonlinear_constraints_gk} into the bootstrap for weakly-coupled $\mathcal{N}=8$ SUSY EFTs, first numerically and then analytically. 

After a brief description of dispersion relations and perturbative bootstrap bounds in Section \ref{sec:bootsetup}, we show in Section \ref{sec:nonlinbootstrap} how to include the nonlinear constraints in the semi-definite optimizer SDPB~\cite{Simmons-Duffin:2015qma} and present examples of the resulting bounds. Excluding infinite spin towers at the mass gap, we find in Section \ref{sec:bifurcation} that positivity and the nonlinear constraints numerically isolate the Virasoro–Shapiro amplitude. In Section \ref{sec:analytic}, we  then show analytically that the Virasoro–Shapiro amplitude is the unique solution to those conditions.

\subsection{Dispersion Relations and Positivity Bounds}\label{sec:bootsetup}

The Wilson coefficients in weakly-coupled theories with $stu$-symmetric amplitudes can be constrained through so-called \emph{positivity bounds} \cite{Tolley:2020gtv,Caron-Huot:2020cmc,Arkani-Hamed:2020blm,Caron-Huot:2021rmr,Caron-Huot:2022ugt,Haring:2023zwu,Albert:2024yap,Berman:2024eid}. These positivity bounds rely on the analytic structure of the amplitudes as a function of complex $s$ with fixed momentum transfer $u$. In particular, the 4-point scalar amplitude $M_4(zz\bar{z}\bar{z})$ is entirely analytic apart from possible simple poles and branch cuts on the real $s$ axis. We assume a cutoff energy scale $\Lambda$ such that the amplitude is also analytic for real $s$ in the range $-\Lambda^2 -u < s < \Lambda^2$.\footnote{The dispersed amplitude is $M_4(zz\bar{z}\bar{z}) = s^4F(s,t,u)$, including the graviton pole of $\kappa^2/stu$. At fixed $u<0$, the $t$-channel pole sits at $s=-u$ and is enclosed by the contour $\mathcal{C}_u$, while due to the $s^4$-factor there is no massless pole in the $s$-channel.} This allows us to relate the low-energy behavior of 
$M_4(zz\bar{z}\bar{z})$, obtained from a small contour $\mathcal{C}_0$ around the origin and $\mathcal{C}_u$ around $s = -u$, to the integral over the non-analyticities on the real $s$ axis 
via a contour-deformation:
\begin{align}
\label{eq:contour}
\frac{1}{2\pi i}\oint_{\mathcal{C}_0+\mathcal{C}_{u}} \frac{M_4(zz\bar{z}\bar{z})}{s^{k+1}} = \int_{\Lambda^2}^{\infty}\frac{ds}{\pi}\left[\frac{1}{s^{k+1}}+\frac{(-1)^{k}}{(s+u)^{k+1}}\right]\text{Im}[M_4(zz\bar{z}\bar{z})]\,.
\end{align}
Here, $k$ defines a number of \emph{subtractions}.
Below we only assume that~\reef{eq:contour} holds for $k \geq 4$. In general, causality and unitarity imply convergence for $k\geq 2$ since the amplitude is bounded in the large-$s$, fixed-$u$ limit as \cite{Caron-Huot:2022ugt,Haring:2022cyf,Bellazzini:2025bay} (technically this holds after \emph{smearing}
against a normalizable wavefunction):
\begin{align}\label{eq:Froissart}    \lim_{|s|\to\infty}\frac{M_4(zz\bar{z}\bar{z})}{s} \leq  \mbox{constant}\, .
\end{align}
The existence of a polynomial upper bound on the growth of the amplitude \reef{eq:Froissart} allows us to use \eqref{eq:contour} to write a \emph{dispersion relation} for any coefficient that grows at least as $s^2$ in the low energy ansatz. Thanks to the overall $s^4$-factor from maximal SUSY, all coefficients $g_n$, $g_n'$, \dots, in the low-energy expansion~\eqref{eq:fstu} of $M_4(zz\bar{z}\bar{z})$ have convergent dispersive expressions~\cite{Haring:2023zwu,Albert:2024yap,Berman:2024eid} that use only $k\geq 4$ subtractions.  Since these project out the graviton pole (which appears only for $k=2$), we analyze \eqref{eq:contour} by expanding both sides around the forward limit. On the left-hand side, we expand the amplitude at low energies and isolate each $g_n$, $g_n'$, \dots, by taking $u$-derivatives and subsequently setting $u=0$. On the right-hand side, we use  
the partial wave expansion 
\begin{align}
    M_4(zz\bar{z}\bar{z}) = 16\pi\sum_{j = 0}^{\infty} (2j+1) \,a_j(s) \,\mathcal{P}_j \!\Big(1+\frac{2u}{s}\Big) \,,
\end{align}
with $a_j(s)$ being the spin-$j$ partial wave of the amplitude and $\mathcal{P}_j(x)$ the spin-$j$ Legendre polynomial. Applying the $u$-derivatives and then setting $u=0$ also on the right-hand side, the result is finally a dispersive representation for each Wilson coefficient $g_n$, $g_n'$, \dots\,. The simplest case is for $g_{2k}$, which are the coefficients of $s^4 (s^2+t^2+u^2)^{k}$ in the amplitude, where we find~\cite{Caron-Huot:2020cmc,Albert:2024yap,Berman:2024eid}
\begin{align}
\label{eq:g2kdrs}
    g_{2k} = \frac{1}{2^{k-1}}\sum_{j=0}^{\infty}\int_{\Lambda^2}^\infty \frac{ds}{s^{2k}} \rho_j(s)\quad \text{for}~~k\geq0\,,
\end{align}
where $\rho_j(s) = 16(2j+1)\text{Im}[a_j(s)]/s^5$ is the spin-$j$ spectral density. The dispersion relations exhibit how the UV spectrum of massive states imprints itself on the low-energy Wilson coefficients.
Importantly, unitarity requires the positivity condition $\rho_j(s) > 0$.

The dispersion relations together with positivity constrain the allowed values of the Wilson coefficients. As a simple example, since all terms in the integrand of \eqref{eq:g2kdrs} are positive and the integration goes over $s > \Lambda^2$, we find
\begin{align}
    0 \,\leq\, 2^{k'} \,g_{2k'}\,\Lambda^{4k'} \,\leq \,2^k \,g_{2k}\,\Lambda^{4k}
    ~~~\text{for}~~k \leq k'\,.
\end{align}
From this equation, it follows that $g_0 = 0$ implies  $g_{2k}=0$. In fact this extends further: if $g_0 = 0$ then all $g_{n}$, $g_{n}'$, \dots, must vanish too. Hence, without loss of generality, we can assume $g_0 > 0$.

Using the dispersive expressions for the Wilson coefficients along with the so-called \emph{null constraints} implied by the crossing symmetry of $f(s,t,u)$, the task of finding numerical upper and lower bounds on the ratio of any Wilson coefficient with respect to $g_0$ can be formulated as a linear optimization problem~\cite{Caron-Huot:2020cmc,Haring:2023zwu,Albert:2024yap,Berman:2024eid}. Here, though, we go beyond these generic bounds by imposing the nonlinear constraints found in Section \ref{sec:Nonlinear_constraints} to further constrain the space of allowed Wilson coefficients.

We briefly comment on the use of the forward limit ($u=0$) to isolate individual Wilson coefficients in \eqref{eq:contour}. In theories with long-range interactions such as gravity, the Taylor expansion of the amplitude around $u=0$ is not strictly well-defined, even for the imaginary part at high energies \cite{Caron-Huot:2024tsk}.
This issue can be avoided by using \emph{smeared} sum rules \cite{Caron-Huot:2021rmr,Caron-Huot:2022ugt,Albert:2024yap,Beadle:2025cdx,Chang:2025cxc,Pasiecznik:2025eqc}, in which the dispersion relations are integrated over some window of momentum transfer, physically to focus on small impact parameters.
However, as long as we neglect low-energy loops and restrict to $k>2$ subtractions (so that the graviton pole is projected out), the long-range interactions are effectively invisible to the formalism and we expect to obtain identical results using either forward limits or smearing techniques.\footnote{Using smeared sum rules with $k\geq 2$, one can find bounds that involve
Newton's constant $\kappa^2$ in addition to
Wilson coefficients. These bounds are infrared safe in $d>4$ dimensions.  We have tested that by imposing the same nonlinear constraints for an EFT in $d=10$ with kinematics restricted to four dimensions, we find the same physical results as without smearing.}

\subsection{Numerical Bootstrap with Nonlinear Constraints}
\label{sec:nonlinbootstrap}

The optimization problem associated with finding bounds on the Wilson coefficients is typically solved using the semi-definite program SDPB~\cite{Simmons-Duffin:2015qma}. However, semi-definite optimization applies only to linear or \textit{convex} quadratic constraints. In order to implement the non-convex constraints~\eqref{eq:nonlinear_constraints_gk} in SDPB, we solve for the coupling constant $\kappa^2$ using the nonlinear constraint for $g_3$, resulting in $\kappa^2 = g_0^2/(2g_3)$, which we then substitute into the remaining relations. Introducing the two variables $\bar{g}_2=g_2/g_0$ and $\bar{g}_3 = g_3/g_0$, the constraints at the lowest orders become
\be
  \label{eq:linSUSYcond}
  \begin{split}
  g_5 - 2 g_2 \, \bar{g}_3 &= 0\,,
  \\
  3 g_6' - 8 g_6 - 2 g_0 \, \bar{g}_3^2 &= 0\,,
  \\
  g_7 -2 g_4 \, \bar{g}_3 - g_0 \, \bar{g}_2^2 \, \bar{g}_3 &= 0\,,
  \\
  g_8' - 8 g_8 -2 g_2 \, \bar{g}_3^2&=0\,, ~~\text{etc}.
  \end{split}
\ee 
For any choice of numerical value for $\bar{g}_3$, most of the constraints are linear in the remaining coefficients and can be readily incorporated into SDPB. This is, however, not the case for the $g_7$ and one of the two $g_9$ constraints. These two cases could be linearized with input of a numerical value also for $\bar{g}_2$, resulting in a scan over the 2d plane $(\bar{g}_2,\bar{g}_3)$. To avoid this, in this section we consider exclusively the constraints that can be implemented by scanning over values of $\bar{g}_3$ only.

\begin{figure}[t]
    \centering
    \includegraphics[width=0.6\textwidth]{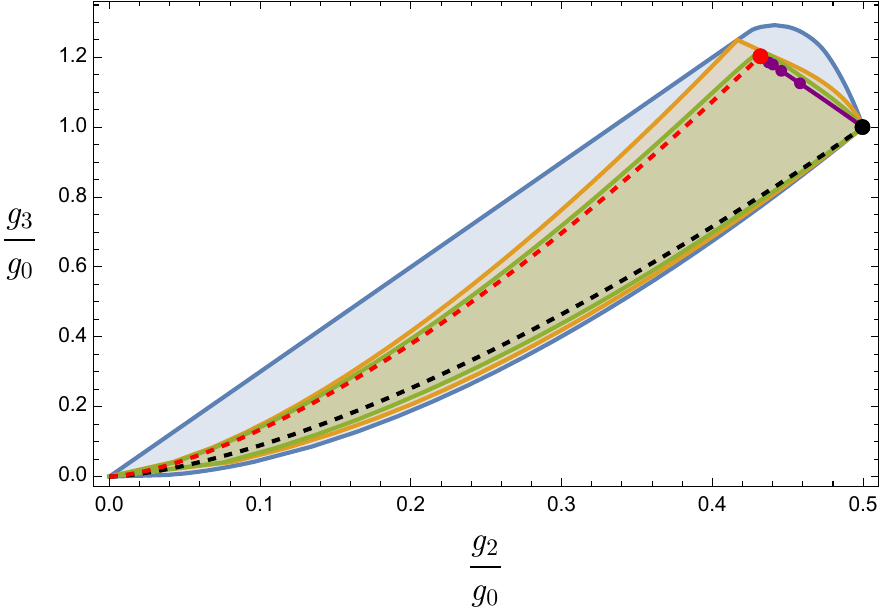}
    \caption{Allowed region in the $(g_2/g_0,g_3/g_0)$ plane. Blue corresponds to the region without nonlinear constraints imposed, yellow includes the first nonlinear constraint in~\eqref{eq:linSUSYcond}, while the green region includes the first and second nonlinear constraints. In red is the scaling curve for the Virasoro–Shapiro amplitude~\eqref{eq:Virasoro-Shapiro}, while in black is that of the IST amplitude~\eqref{eq:M4IST}. The purple line corresponds to the coefficients of \eqref{eq:Wilson_coeffs_MN} for $m^2 = \Lambda^2$, with the purple dots marking the cases of $N = 2,3,4,5$, with increasing $N$ getting closer to the string point.}
    \label{fig:numericalboundsNG}
\end{figure}

In Figure \ref{fig:numericalboundsNG}, we show the allowed region in the $(g_2/g_0,g_3/g_0)$ plane when different sets of nonlinear constraints are imposed. The general allowed region without any nonlinear constraints is shown in blue. In yellow, we include only the first constraint in \eqref{eq:linSUSYcond}, while green corresponds to including the first two constraints in \eqref{eq:linSUSYcond}. 

The Virasoro–Shapiro amplitude~\eqref{eq:Virasoro-Shapiro} is shown as a dashed red line which corresponds to the \emph{scaling curve} for the string amplitude: values of the coefficients for different choices of $\alpha'\Lambda^2\leq 1$, with the red dot being $\alpha'\Lambda^2 = 1$. Similarly, the black dashed line shows the IST amplitude~\eqref{eq:M4IST} for different $\Lambda^2/m^2 \leq 1$, and the black dot being $\Lambda^2/m^2 =1$. Without the nonlinear constraints, the string amplitude lives on the interior of this projection, with the $\alpha'\Lambda^2 = 1$ point sitting somewhat close to the boundary, but not at any obviously special point. The IST, on the other hand, is a corner in the allowed region and the associated scaling curve appears to be close to the lower boundary.\footnote{In \cite{Berman:2024eid}, it was shown that as increasingly many null constraints are imposed, the boundary moves closer to the IST scaling curve, suggesting that it defines the true boundary.}
When we introduce the nonlinear constraints, however, we find that a second corner appears. As an increasing number of the nonlinear constraints are imposed, this corner moves towards the string amplitude \reef{eq:Virasoro-Shapiro} with $\alpha'\Lambda^2 = 1$! Similarly, the allowed region changes such that the scaling curve of the string amplitude lies close to the upper left boundary. Therefore, the nonlinear constraints clearly select both the IST and the string amplitude as special points in the bootstrap space.

The parameter space outside of the colored regions in Figure \ref{fig:numericalboundsNG} is rigorously excluded. The bounds illustrate how incorporating the nonlinear constraints results in a non-convex region, but since the plots are made with only a relatively small subset of the crossing null constraints and nonlinear constraints, the bounds are not fully converged. Including higher orders in the null constraints and nonlinear constraints is thus expected to further shrink the allowed region. Yet, even from these low-order constraints, it looks plausible that the bounds converge to the region bounded by the scaling curves of the Virasoro–Shapiro amplitude, the IST amplitude, and the purple line connecting the red and black dots in Figure \ref{fig:numericalboundsNG}, which corresponds to the coefficients of \eqref{eq:Wilson_coeffs_MN} for $m^2 = \Lambda^2$.

With just positivity of the spectral density imposed, any sum of positive spectral densities is also allowed, so these projective spaces are always convex. With the non-convex constraints of \eqref{eq:linSUSYcond}, though, the sum of two allowed spectral densities is no longer allowed and the regions can be non-convex. Importantly, this means that the allowed region between the string and the IST cannot be simply linear sums of the two amplitudes.

\subsection{Bifurcation from a Mass Gap}
\label{sec:bifurcation}

Similarly to the exclusion plot in Figure \ref{fig:numericalboundsNG}, a non-convex region with two corners, one at the IST and one at the open string, was observed in \cite{Berman:2025owb}, when considering the nonlinear constraints arising from the so-called hidden zero splittings~\cite{Arkani-Hamed:2023swr}. In that case, it turned out that the open string was the unique amplitude once the IST was explicitly excluded by allowing only a finite number of states to be exchanged at $s = \Lambda^2$. Doing so ruled out not just the pure IST, but also the entire region connecting the two points.

\begin{figure}[t]
    \centering
    \includegraphics[width=0.49\textwidth]{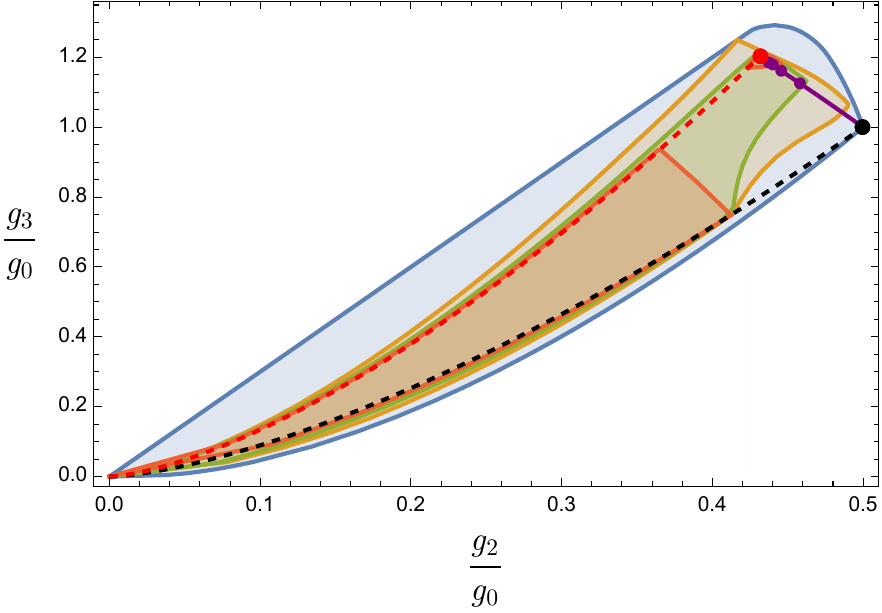}
    \hfill
    \includegraphics[width=0.49\textwidth]{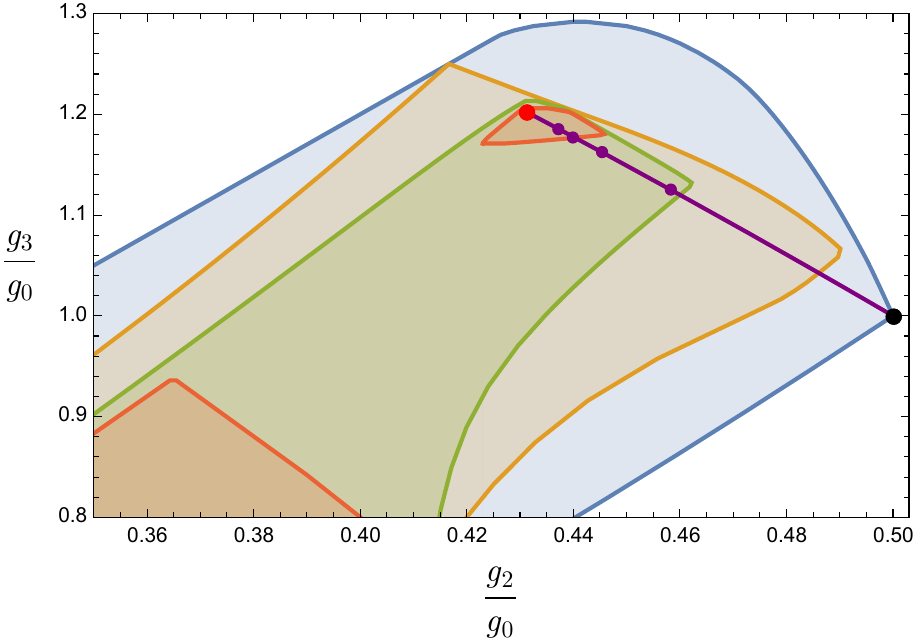}
    \caption{
    Left: Full allowed region in the $(g_2/g_0,g_3/g_0)$ plane. Blue corresponds to the region without nonlinear constraints imposed, yellow includes the first nonlinear constraint in~\eqref{eq:linSUSYcond} with a gap $\mu_c = 1.1$, and green includes both the first and second nonlinear constraints with the same gap $\mu_c = 1.1$. The orange region has $\mu_c = 1.1$ and includes the first, second, and fourth constraints in~\eqref{eq:linSUSYcond}, ignoring the third constraint that additionally requires us to fix $\bar{g}_2$. The red and black curves provide the scaling curves for the Virasoro–Shapiro amplitude~\eqref{eq:Virasoro-Shapiro} and the IST amplitude~\eqref{eq:M4IST}, respectively. Right: Zoomed-in view near the diagonal connecting the string with $\alpha'\Lambda^2 = 1$ and the IST with $\Lambda^2/m^2 = 1$. As in Figure \ref{fig:numericalboundsNG}, the purple line in both plots corresponds to the coefficients of \eqref{eq:Wilson_coeffs_MN} for $m^2 = \Lambda^2$, with the purple dots marking the cases of $N = 2,3,4,5$, with increasing $N$ getting closer to the string point.}
    \label{fig:numericalbounds}
\end{figure}

We consider the same approach here: allow states only up to some finite spin to be exchanged at $s = \Lambda^2$, but be entirely agnostic about the spectrum above some scale $\mu_c \Lambda^2$ for $\mu_c>1$. With the Regge limit behavior assumed in \eqref{eq:Froissart}, requiring only finitely many spins at the lowest mass immediately implies that only scalars can be exchanged at $s=\Lambda^2$ \cite{Berman:2024owc}.  With this assumption, the spectral density $\rho_j(s)$ can be written as
\begin{align}
    \rho_j(s) = \frac{|\lambda_0|^2}{\Lambda^{10}}\delta\left(1-\frac{s}{\Lambda^2}\right)\delta_{j,0} + \Theta\left(\frac{s}{\mu_c\Lambda^2}-1\right)\tilde{\rho}_j(s)\,,
\end{align}
where $|\lambda_0|^2$ is the (dimensionless) squared coupling of the massless states to the possible scalar at $s = \Lambda^2$ and $\Theta(x)$ is the Heaviside function, so that $\tilde{\rho}_j(s)$ denotes the spectral density at $s \geq \mu_c\Lambda^2$. Substituting in~\eqref{eq:g2kdrs}, the dispersion relations for each of the $g_{2k}$ coefficients become
\begin{align}
    g_{2k} = \frac{1}{2^{k-1}}\left(\frac{|\lambda_0|^2}{\Lambda^{4(k+2)}} + \sum_{j=0}^{\infty}\int_{\mu_c\Lambda^2}^{\infty}\frac{ds}{s^{2k}}\tilde{\rho}_j(s)\right),
\end{align}
reflecting that we allow for arbitrary unitarity contributions above $\mu_c\Lambda^2$. 

To implement this additional cutoff in practice in the numerical bootstrap, we must make some choice of $\mu_c$. The result is qualitatively independent of the cutoff so long as $1< \mu_c \leq 2$. In particular it does not depend on taking $\mu_c$ close to the string-inspired choice of $\mu_c = 2$. To illustrate this, we choose $\mu_c = 1.1$ in the plots shown in Figure \ref{fig:numericalbounds}. Analogously to \cite{Berman:2025owb}, once we require finitely many spins at the mass gap, the region becomes increasingly concave as we impose more nonlinear constraints. Ultimately, the allowed region bifurcates into a small island of parameter space around the string and a larger region in which $|\lambda_0|^2 = 0$. 

Therefore, any amplitude on the line between the string and the IST in Figure \ref{fig:numericalboundsNG}, other than the string itself, must also have states of arbitrarily large spin exchanged at $s = \Lambda^2$. As discussed above, these theories are not simply sums of the string amplitude with the IST, so they must be some distinct amplitudes with at least one spin tower.

The numerical results indicate that the closed string 4-point amplitude is isolated as the unique tree-level amplitude completion compatible with $\mathcal{N}=8$ SUSY, tree-level factorization in the EFT, and peculiar parity. We now derive this result analytically.

\subsection{Analytic Bootstrap: Virasoro–Shapiro and New Infinite Spin Towers}
\label{sec:analytic}

By imposing the nonlinear constraints \eqref{eq:nonlinear_constraints_gk} on the low-energy ansatz \eqref{eq:fstu}, we find evidence that the EFT amplitude can be resummed into an exponential form. We have verified this up to $\mathcal{O}(s^{13})$, the highest order we have checked. Following similar logic to \cite{Wan:2026pjq}, upon assuming that this exponentiated form holds to all orders, we prove the following results: that any maximally SUSY gravitational amplitude that satisfies the nonlinear constraints must admit a meromorphic product form and that it must coincide with the Virasoro–Shapiro amplitude if it has finitely many spins at the lowest mass level.

We begin by stating the observed exponentiated form of the nonlinear constraints.  Up to $\mathcal{O}(s^{13})$, the only coefficients in \eqref{eq:fstu} not fixed by the nonlinear constraints \eqref{eq:nonlinear_constraints_gk} are the coefficients of $(s^2+t^2+u^2)^{k}$ for each integer $k$, which we label $g_{2k}$. The remaining coefficients are all fixed such that the low-energy expansion of the 4-point amplitude matches an exponential:
\begin{align}
\label{eq:expamp}
M_4(zz\bar{z}\bar{z}) =&\, s^4 \left[ \frac{\kappa^2}{stu} + g_{0} + g_{2} \, (s^2+t^2+u^2) + \frac{g_{0}^2}{2 \kappa^2} \, stu + g_{4} \, (s^2+t^2+u^2)^2 + \ldots  \right] \nonumber \\
=&\, \frac{\kappa^2s^4}{stu}\exp\left(\sum_{k=0}\frac{2^{k}g_{2k}}{\kappa^2(2k+3)}(s^{2k+3}+t^{2k+3}+u^{2k+3})\right).
\end{align}
This form holds up to at least $\mathcal{O}(s^{13})$ and has passed nontrivial checks at $\mathcal{O}(s^{10})$, $\mathcal{O}(s^{12})$, and $\mathcal{O}(s^{13})$ where there exist multiple $stu$-symmetric polynomials. Note that \reef{eq:expamp} is a generalization of the exponentiated form of the Virasoro–Shapiro amplitude~\reef{eq:Virasoro-Shapiro-exp}, which is obtained from \reef{eq:expamp} when restricting the Wilson coefficients $g_{2k}$ to the $\zeta$-values in \eqref{eq:g2m_VS}.

We emphasize that the sum in the exponential of \eqref{eq:expamp} involves only the subset $g_{2k}$ of Wilson coefficients which do not vanish in the forward limit. The agreement between the two formulas becomes trivial in this limit since the exponent vanishes term-by-term and the exponential collapses to a sum,
\begin{align}  \frac{s^{2k+3}+t^{2k+3}+u^{2k+3}}{(2k+3)} ~~\xrightarrow{u \to 0}~~ s t u \times s^{2k}
    =\mathcal{O}(u)\,.
\end{align}
Thus the nonlinear constraints reconstruct the amplitude from its forward limit.

To involve the positivity conditions, 
we substitute the dispersive 
representation of the $g_{2k}$ coefficients~\eqref{eq:g2kdrs} into the exponent of \reef{eq:expamp} to obtain (for ease of notation, we drop the argument from $M_4(zz\bar{z}\bar{z})$ in the rest of this section):
\begin{equation}
   M_4 =
   \frac{\kappa^2s^4}{stu}
\exp\left(
\frac{2}{\kappa^2}\sum_{k=0}\int_{\Lambda^2}^\infty ds' \rho(s')  \frac{(s')^{-2k}}{2k+3}(s^{2k+3}+t^{2k+3}+u^{2k+3})
\right)\,,
\end{equation}
where we define $\rho(s) \equiv \sum_j\rho_j(s)$. Interchanging the sum and integral and defining the rescaled spectral density $\tilde{\rho}(s')=\frac{s'^3}{\kappa^2}\rho(s')$, the general solution to the nonlinear constraints may thus be expressed as
\begin{equation} \label{M4 nonlinear exp}
   M_4 
= \frac{\kappa^2s^4}{stu}\exp\left(
\int_{\Lambda^2}^\infty ds'\tilde{\rho}(s') \log\left(\frac{(s'+s)(s'+t)(s'+u)}{(s'-s)(s'-t)(s'-u)}\right)
\right).
\end{equation}
The two-variable function $M_4(s,t)$ is now entirely parameterized by $\tilde\rho(s)$.

\paragraph{Meromorphy.}
The form \eqref{M4 nonlinear exp}  correlates poles at positive $s$ with zeros at negative $s$ and other Mandelstam invariants.  In fact, for any nonvanishing spectral density, the branch cut of $\log(s'+t\pm i0)$ will cause the exponent to be non-analytic in $t$ for some $t<0$, where the amplitude should be analytic.
The only way to reconcile these statements is if the discontinuity of the exponent is always an integer multiple of $2\pi$, which requires that $\tilde{\rho}(s')$ be a sum of localized terms $c_n\delta(s'-m_n^2)$ with integer coefficients $c_n$.

Since any coefficient $c_n>1$ would lead to a power singularity $(m_n^2-s)^{-c_n}$ in the amplitude $M_4$, which would be incompatible with residues having a positive partial wave expansion, we conclude that all nonvanishing $c_n$'s must be equal to 1.
Hence the amplitude must be expressible as a product over a discrete list of masses $\{m\}$:
\begin{align} \label{M4 nonlinear prod}
    M_4^{\{m\}} = \frac{\kappa^2s^4}{stu}
    \, \prod_n \frac{(m_n^2+s)(m_n^2+t)(m_n^2+u)}{(m_n^2-s)(m_n^2-t)(m_n^2-u)}\,.
\end{align}
All we have used so far are the nonlinear constraints, forward dispersion relations, analyticity for negative $t$, and a mild form of positivity; the list of masses can be either finite or infinite, but every $m_n$ must be distinct.

A comment is in order regarding analyticity in $t$. In a gapped theory, ${\rm Im}\,M_4(s,t)$ at fixed $s$ is guaranteed to be analytic for $\cos\theta$ in the so-called small Lehmann ellipse, which contains the physical region $|\cos\theta|<1$ \cite{Correia:2020xtr}; this is deduced from absolute convergence of positive sums of Legendre polynomials.
In the massless limit, the ellipse degenerates to the real interval $-s<t<0$ and we are not aware of a rigorous axiomatic treatment which would forbid singularities in this interval.
However, we still consider this possibility as pathological since we would have no idea how to interpret and handle such a singularity. Indeed, unlike other familiar singularities that appear in scattering amplitudes, it would be impossible to make sense of it by a standard shift $t\pm i0$ since this would break the symmetry between $t$ and $u=-s-t\mp i0$.
For this reason we accept real analyticity in the physical region $-s<t<0$ as a basic axiom of massless scattering theory.

The constraints are not yet exhausted and we must still require that each residue in \eqref{M4 nonlinear prod} admits a positive partial wave decomposition.  The residue at the $n$th mass level is
\begin{equation} \label{residue i}
    -{\rm Res}_{s=m_n^2}
    M_4^{\{m\}} = \frac{2\kappa^2 m_n^8}{(m_n^2-t)(2m_n^2+t)}
\prod_{j\neq n} \frac{(m_j^2+m_n^2)(m_j^2+t)(m_j^2-m_n^2-t)}{(m_j^2-m_n^2)(m_j^2-t)(m_j^2+m_n^2+t)}\,.
\end{equation}
Although a general analysis of the partial wave decomposition of this residue is beyond our scope, we analyze some special cases.

\paragraph{Uniqueness of Virasoro–Shapiro for Finite Spins.}

Consider the lowest mass level $m_1$. If only finitely many spins exist at that level, the residue \eqref{residue i} must be a polynomial in $t$.
Inspection of the formula reveals that the pole at $t=m_1^2$ can only be canceled if there exists a second particle with $m_2^2-m_1^2=m_1^2$, i.e.~$m_2^2=2m_1^2$.

The contribution from $m_2$ to \eqref{residue i} produces a new pole at $t=2m_1^2$, which can only be canceled by a third particle with $m_3^2=3m_1^2$, and so on. Hence the structure \eqref{M4 nonlinear prod} can only accommodate a polynomial residue if the spectrum is exactly linear ($m_n^2=n m_1^2$):
\begin{equation}
    M_4^{\text{finite-spins}}
    = \frac{\kappa^2s^4}{stu}
    \, \prod_{n=1}^\infty \frac{(nm_1^2+s)(nm_1^2+t)(n m_1^2+u)}{(nm_1^2-s)(nm_1^2-t)(n m_1^2-u)}\,,
\end{equation}
which coincides precisely with the product form of the Virasoro–Shapiro amplitude \eqref{eq:product_form_VS} with $\alpha'=m_1^{-2}$ !

\paragraph{Finite-Product Amplitudes.}

For other choices of masses, positivity of all partial waves is nontrivial. A simple example is a finite product~\eqref{M4 nonlinear prod} with two masses $m_1<m_2$. Projecting the $m_2$ residue onto the scalar exchange we get the coupling
\begin{align}
\label{eq:couplingpositivity}
    |\lambda_{0,m_2}|^2 = -\frac{4 r^2 (r+1) \left(-6 r^2 \log (r+1)+(r (2 r+5)+1) r \log 2-\log 4\right)}{3 (r-1)^2 (r+2) (2 r+1)}\,,
\end{align}
where $r = m_2^2/m_1^2$ is the ratio of the tower masses. This is only positive for $1.25 < r < 5.63$, where the numbers are determined as the roots of the numerator of (\ref{eq:couplingpositivity}). However, this test is not enough on its own. By studying the partial waves of the states at $m_1$,  one can show that positivity at large spin requires additionally $r \geq 2$.
We find no other constraints, hence the two-mass product amplitude is unitary if
\begin{align}
    2 \leq \frac{m_2^2}{m_1^2} \leq 5.63\,.
\end{align}
The lower bound is realized by the amplitude $M_4^{(2)}$ defined in \eqref{eq:M4N}, which suggests that its Wilson coefficients live at the boundary of the space allowed by positivity and the nonlinear constraints.

For products involving more than two masses, the criterion for determining the allowed masses becomes more complicated.
However, we observe that if the second mass takes its minimal allowed value, $m_2^2 = 2m_1^2$, then the lower bound on $m_3^2$ while retaining positivity is $m_3^2 = 3m_1^2$, again exactly matching $M_4^{(3)}$, and so on for any $M_4^{(N)}$ with integer $N$. This suggests that the amplitudes $M_4^{(N)}$ in \reef{eq:M4N} are in some sense ``extremal'' and live on the boundary of the allowed region. Indeed, as we increase the number of nonlinear and numerical constraints in Figure \ref{fig:numericalboundsNG}, the upper right boundary appears to converge to the line corresponding to the $M_4^{(N)}$ amplitudes with $m_1^2=1$.

\section{Conclusions and Outlook}
\label{sec:conclusions}

In this paper, we have studied the low-energy effective field theory of four-dimensional $\mathcal{N}=8$ supergravity, deformed by a general tower of higher-derivative corrections. Our analysis is based on on-shell superamplitudes through 6 points, whose low-energy expansion coefficients encode the Wilson coefficients of the higher-derivative operators. We impose maximal $\mathcal{N}=8$ supersymmetry, $\text{SU}(4)\times\text{SU}(4)$ R-symmetry, consistent tree-level factorization, and peculiar parity. The latter requires a particular scalar subsector to be parity invariant at tree level, even though this invariance is necessarily violated by fermion loops. Notably, we show that these assumptions are sufficient to isolate the weakly-coupled closed string amplitude. Related questions have been investigated in~\cite{Caron-Huot:2016icg,Huang:2020nqy,Chiang:2023quf,Arkani-Hamed:2023jwn,
Cheung:2024uhn,Berman:2024wyt,Cheung:2024obl,Bhat:2024agd,Berman:2025owb,
Cheung:2025tbr,Eckner:2025kve,Elvang:2026pmc,Basile:2026gnd,Wan:2026pjq,Xu:2026kix}, which also consider deriving the closed string amplitude from some minimal set of assumptions. A distinguishing feature of our analysis is that, despite their strength, the assumptions we impose are not string-theoretic in nature.

Since peculiar parity is an approximate symmetry that is not well-defined on the full Hilbert space, it is neither logically implied by the other symmetry assumptions we have made nor obviously necessary for the theory's internal consistency. Nevertheless, it is remarkable that it provides a ``symmetry-based'' explanation for the meromorphy and product structure of string amplitudes (see Subsection \ref{sec:analytic}).

More generally, our analysis suggests that higher-point amplitudes can sharply constrain the lower-point data of an effective field theory in the presence of symmetries or properties that cannot be made manifest at the level of the Lagrangian, such as maximal supersymmetry. Such ``hidden'' structures are important precisely because their consequences may be invisible in any fixed-multiplicity operator basis. The nonlinear relations among the 4-point Wilson coefficients are invisible at 4 points and emerge only through factorization when a property of the 6-point amplitude is imposed. Although this possibility was already suggested by earlier work~\cite{Berman:2025owb,Elvang:2026pmc}, our results show that the same mechanism operates in a gravitational theory. An important open question is whether analogous constraints persist in some form in theories with substantially less symmetry.

These observations suggest several directions for future work. One is to search for related phenomenological applications. For example, an analog of peculiar parity constrains contributions to the $W^+W^-Z$ vertex,
which contains a $C$- and $P$-odd but $CP$-even form factor $g_5^Z$~\cite{Hagiwara:1986vm}. Its predicted smallness in the Standard Model means that experimental bounds on $g_5^Z$ constrain beyond-the-Standard Model scenarios with additional sources of $C$ and $P$ violation that conserve $CP$. It would be valuable to identify other peculiar symmetries, including generalized~\cite{Gaiotto:2014kfa} and non-invertible~\cite{Chang:2018iay} analogs, which may impose novel constraints on extended operators and their selection rules.

A second direction is to extend the analysis to curved spacetime, in particular to AdS/CFT \cite{Maldacena:1997re,Gubser:1998bc,Witten:1998qj}. Boundary correlators of supersymmetric theories in anti-de Sitter space should obey linearized SUSY Ward identities \cite{Belitsky:2014zha}, and our analysis relies only on supersymmetry, locality, and factorization, each of which should have a counterpart, in some form, for such correlators. It would be interesting to ask whether large-$N$ conformal field theories possess approximate symmetries similar to peculiar parity, and whether these can constrain finite 't~Hooft coupling corrections to higher-point correlation functions. If so, the strategy developed here could help constrain holographic correlators away from the supergravity limit.

A third direction concerns the role of the R-symmetry itself. Imposing $\text{SU}(4)\times\text{SU}(4)$ R-symmetry together with peculiar parity singles out the closed superstring, raising the question of whether other R-symmetry groups, either alone or supplemented by an analog of peculiar parity, can similarly isolate unique nontrivial completions of tree-level $\mathcal{N}=8$ supergravity. An analogous analysis of M-theory, for example, might reveal such a distinguished subgroup. Classifying such symmetry structures and the theories they select would clarify whether the closed superstring belongs to a broader family of uniquely determined completions or whether the $\text{SU}(4)\times\text{SU}(4)$ case is genuinely exceptional.

\section*{Acknowledgments}

The authors thank Nima Arkani-Hamed, Carolina Figueiredo, Hare Krishna, Nathan Seiberg, and Edward Witten for useful discussions. 
JB is supported by a Predoctoral Fellowship from the Rackham Graduate School and a Graduate Summer Fellowship from the Leinweber Institute for Theoretical Physics at the University of Michigan.
The work of SCH is supported by the National Science and Engineering Council of Canada (NSERC), funding reference SAPIN/00028-2022, and the Canada Research Chair program, reference number CRC-2022-00421. The work of AC, HE, and RM is supported in part by Department of Energy grant DE-SC0007859. AC was supported in part by Cottrell SEED Award
number CS-SEED-2023-004 from the Research Corporation for Science Advancement. RM is also funded by the Leinweber Postdoctoral Fellowship from the University of Michigan. AH is grateful to the Simons Foundation as well as the Edward and Kiyomi Baird Founders’ Circle Member Recognition for their support. LLL is supported by a Graduate Research Fellowship from the National Science Foundation under grant DGE-2241144 and a Rackham Merit Fellowship from the University of Michigan.
This paper was finished while HE was at the Aspen Center for Physics, which is supported by National Science Foundation grant PHY-2210452.

\appendix

\section{Details on the 5-point MHV Amplitude}
\label{app:5pt_analysis}

In this appendix, we provide further details on the 5-point MHV amplitude $M_5(--+++)$ from Section~\ref{sec:5pt_amps}. Concretely, we fix the majority of the parameters in its ansatz by imposing consistent factorization to lower-point amplitudes in the physical channels.

First, recall from~\eqref{eq:5pt_MHV_superamp} and~\eqref{eq:5ptansatz} that the 5-point MHV superamplitude can be written as
\begin{equation}
  \mathcal{M}^{\text{MHV}}_5(12345) = \frac{-G_5(12345)}{\ab{12}\ab{13}\ab{14}\ab{15}\ab{23}\ab{24}\ab{25}\ab{34}\ab{35}\ab{45}} \, \delta^{(16)}(\widetilde{Q}) \,.
\end{equation}
Importantly, the superamplitude is fully permutation symmetric in the labels of the particles. Since the denominator on the right-hand side is antisymmetric under the exchange of any pair of particles, and the delta function is invariant, the polynomial $G_5(12345)$ must therefore be fully permutation antisymmetric. Based on the definition of $G_5(12345)$ in~\eqref{eq:def_G5}, we see that the polynomial $V_5(12345)$ is antisymmetric too. On the other hand, since the Levi–Civita contraction $\epsilon[1234]$ is already fully antisymmetric, $Q_5(12345)$ must be instead symmetric under all permutations of the momenta. Enforcing then the respective symmetries on the polynomials $V_5$ and $Q_5$, we find that the number of coefficients $v_{k,r}$ and $q_{k,r}$ is greatly reduced compared to the $\binom{k+4}{4}$ terms expected for a generic polynomial at order $s^k$; see the counting in the first row of Table~\ref{tab:5pt_ansatz}.

\paragraph{Factorization Constraints.} In order to fix some of the coefficients $v_{k,r}$ and $q_{k,r}$ in the ansatz of the 5-point MHV amplitude, we can demand the correct factorization behavior on the physical poles. To begin with, we have the following factorization on the $s_{15}$ pole:
\begin{align}
\label{eq:M5graviton_factorization51}
  M_5(--+++)\Big|_{s_{15}\,\text{pole}}
  =&\, - \ \begin{tikzpicture}[baseline={([yshift=-0.1cm]current bounding box.center)}] 
	\node[] (a) at (0,0) {};
    \node[] (a1) at ($(a)+(-0.75,0.75)$) {};
    \node[] (a3) at ($(a)+(-0.75,-0.75)$) {};
	\node[] (b) at ($(a)+(3,0)$) {};
    \node[] (b1) at ($(b)+(0.75,0.75)$) {};
    \node[] (b2) at ($(b)+(1,0)$) {};
    \node[] (b3) at ($(b)+(0.75,-0.75)$) {};
	\draw[line width=0.2mm,decoration={snake, amplitude=1.25mm, segment length=2.5mm, pre length=.5mm, post length=.5mm},decorate,double,double distance=0.2ex] (a1.center) -- (a.center);
    \node[] at ($(a1)+(-0.2,0.2)$) {$5^+$};
    \draw[line width=0.2mm,decoration={snake, amplitude=1.25mm, segment length=2.5mm, pre length=.5mm, post length=.5mm},decorate,double,double distance=0.2ex] (a3.center) -- (a.center);
    \node[] at ($(a3)+(-0.3,-0.2)$) {$1^-$};
    \draw[line width=0.2mm,decoration={snake, amplitude=1.25mm, segment length=2.5mm, pre length=.5mm, post length=.5mm},decorate,double,double distance=0.2ex] ($(a)+(0,-0.05)$) -- ($(b)+(0,-0.05)$);
    \draw[line width=0.2mm,decoration={snake, amplitude=1.25mm, segment length=2.5mm, pre length=.1mm, post length=0mm},decorate,double,double distance=0.2ex] (b.center) -- (b1.center);
    \node[] at ($(b1)+(0.3,0.2)$) {$4^+$};
    \draw[line width=0.2mm,decoration={snake, amplitude=1.25mm, segment length=2.5mm, pre length=0.1mm, post length=0mm},decorate,double,double distance=0.2ex] ($(b)+(0.05,-0.05)$) -- ($(b2)+(0.25,-0.05)$);
    \node[] at ($(b2)+(0.55,0)$) {$3^+$};
    \draw[line width=0.2mm,decoration={snake, amplitude=1.25mm, segment length=2.5mm, pre length=.1mm, post length=0mm},decorate,double,double distance=0.2ex] ($(b)+(-0.15,-0.08)$) -- ($(b3)+(-0.13,-0.1)$);
    \node[] at ($(b3)+(0.2,-0.2)$) {$2^-$};
    \node[] at ($(a)+(0.95,-0.45)$) {$(-P)^{+}$};
    \node[] at ($(b)+(-0.7,-0.425)$) {$P^{-}$};
    \filldraw[gray!25] (a.center) circle (0.35cm);
    \draw[line width=0.2mm] (a.center) circle (0.35cm);
    \filldraw[gray!25] (b.center) circle (0.35cm);
    \draw[line width=0.2mm] (b.center) circle (0.35cm);
    \node[] at ($(a)+(0,0)$) {$M_3$};
    \node[] at ($(b)+(0,0)$) {$M_4$};
\end{tikzpicture} \nonumber \\
    =&\, - M_3(5^+ \, 1^- \, (-P)^+) \, \frac{1}{s_{15}} \, M_4(P^- \, 2^- \, 3^+ \, 4^+) \nonumber \\
    =&\,- \kappa \, \frac{\langle 12 \rangle^6 [15] [34]^4}{\langle 15 \rangle \langle 25\rangle^2} F(s_{23}, s_{24}, s_{34})\,.
\end{align}
Here, we use $P = P_{15}$ and the analytic continuation convention $|-p\> = -|p\>$, $|-p] = |p]$. Moreover, to obtain the final result, we have simplified the spinor brackets using identities such as $[5P] \<P2\> = [51]\<12\>$.

The procedure now is to match this factorization with our ansatz from \eqref{eq:5ptansatz}--\eqref{eq:def_G5} order by order, beginning with the leading order at $s^1$. First, note that $V_5(12345)$ vanishes at leading order due to the permutation antisymmetry that it must satisfy. Thus, we simply find
\begin{equation}
    q_{0, 1} = \kappa^3\,,
\end{equation}
which is easy to see comparing the ansatz to the leading order in~\eqref{eq:LO5ptamp}. At the next order, which corresponds to $\mathcal{O}(s^2)$, the permutation-symmetric polynomial $Q_5(12345)$ vanishes due to momentum conservation. Then, matching to the residue in~\eqref{eq:M5graviton_factorization51}, we find that the only free coefficient vanishes,
\begin{equation}
    v_{3, 1} = 0\,.
\end{equation}
Similarly, at order $s^3$ the only coefficient $q_{2, 1}$ also vanishes. Instead, at order $s^4$, two of the parameters are fixed in terms of the 4-pt Wilson coefficient $g_{0}$, while the third one vanishes. A summary of our results up to $\mathcal{O}(s^{12})$ is provided in Table~\ref{tab:5pt_ansatz}. As can be seen, some coefficients remain free starting at order $\mathcal{O}(s^9)$. Notably, performing the matching procedure on the remaining 2-particle factorization channels poses no further constraints on these free coefficients.

\begin{table}[t]
\centering
\vspace{4pt}
\begin{tabular}{|l|cccccccccccc|}
\hline
\rule{0pt}{0.8\normalbaselineskip}
\!\!Order in $M_5(--+++)$ & $s^1$ & $s^2$ & $s^3$ & $s^4$ & $s^5$ & $s^6$ & $s^7$ & $s^8$  & $s^9$ & $s^{10}$ & $s^{11}$ & $s^{12}$ \\
\hline
\rule{0pt}{0.8\normalbaselineskip}
\!\!Ansatz & 1 & 1 & 1 & 3 & 4 & 5 &  9 &  12 & 16 & 22 & 29 & 36 \\[0.05cm]
$s_{15}$ pole residue matching & 0 & 0 & 0 & 0 & 0 & 0 & 0 & 0 & 1 & 1 & 3 & 4 \\
\hline
\end{tabular}%
\caption{Number of free parameters (other than the 4-point coefficients $g_{k}$) in the ansatz of the 5-point MHV amplitude $M_5(--+++)$ at each order up to $\mathcal{O}(s^{12})$, and the resulting simplification after matching the residue at the $s_{15}$ pole with the physical factorization channel.
}
\label{tab:5pt_ansatz}
\end{table}

\paragraph{Local contact terms.} In general, we expect the coefficients that remain free after matching to the physical factorization residues to correspond to local contact terms. As a sanity check, we can actually construct these local contact terms explicitly using the spinor-helicity formalism. To do so, we first note that pure graviton MHV local matrix elements $\langle \dots \, i^{-} \, \dots \, j^{-} \, \dots \, \rangle $ must satisfy the relation
    \begin{equation}
        \langle \dots \, i^{-} \, \dots \, j^{-} \, \dots \, \rangle  = \frac{\langle i j \rangle^8}{\langle 12 \rangle^8} \, \langle --+ \, \dots \, +\rangle\,.
    \end{equation}
Since they are local contact terms, they cannot have any poles. Thus, at 5 points we find that
    \begin{equation}
        \langle --+++\rangle \propto \langle 12\rangle^8 B[+++++] \,,
    \end{equation}
where the function $B[+++++]$ is fully permutation symmetric. Moreover, it must have correct little group scaling, meaning that each spinor square bracket $| k ]$ appears exactly 4 times. Therefore, the lowest possible order at which the first local contact term can be constructed is $\mathcal{O}(s^9)$. In particular, there is only one such local contact term, namely
\begin{equation}
\langle - - +++\rangle = \langle 12 \rangle^8 ( [12]^2 [15]^2 [2 5]^2 [3 4]^4 + \text{perms})\,.
\end{equation}
This expression precisely matches the prefactor of the free coefficient found at $\mathcal{O}(s^9)$ in the 5-pt EFT amplitude. Similarly, we have verified that the prefactors of the free coefficients in the EFT amplitude exactly match the expected local matrix elements up to order $\mathcal{O}(s^{12})$.

\section{Details on Other 6-point Helicity Sectors}
\label{app:6ptHelSect}

In this appendix, we gather the details of the 6-point non-NMHV superamplitudes, i.e. the MHV, N$^{(1,0)}$MHV and N$^{(2,0)}$MHV helicity sectors.

\subsection{MHV Amplitudes}
\label{sec:6pt_MHV}

We begin with the 6-point MHV sector, with a superamplitude that can be written as
\be
  \mathcal{M}^{\text{MHV}}_6(123456)
   = \frac{M_6(--+++\,+)}{\<12\>^8} \, 
  \delta^{(16)}\big( \widetilde{Q}\big) \, .
\ee
Similarly to the 5-point MHV amplitude from~\eqref{eq:5ptansatz}, we can construct a simple ansatz for the 6-point MHV amplitude,
\begin{equation}
\label{eq:ansatz_6pt_MHV}
M_{6}(--+++\,+)=M^{\text{LO}}_6(--+++\,+) \, \left( V_{6}(123456)+\sum_{i=1}^5\epsilon_i\,Q_{6,i}(123456) \right) ,
\end{equation}
where $M^{\text{LO}}_6(--+++\,+)$ denotes the leading-order SUGRA amplitude, which can be computed e.g.~using BCFW recursion~\cite{Britto:2004ap,Britto:2005fq}. In the equation, $V_6$ and $Q_{6,i}$ denote arbitrary polynomials in the nine 6-point Mandelstam variables from~\eqref{eq:6pt_Mandelstams}. Note that, compared to the 5-point MHV ansatz from~\eqref{eq:5ptansatz}--\eqref{eq:def_G5}, which only features a single Levi–Civita, at 6 points we can have five independent Levi–Civita contractions $\epsilon_i$, given in~\eqref{eq:6pt_LC}. Also, just like for the 5-point MHV amplitude discussed in Appendix~\ref{app:5pt_analysis}, in principle it is straightforward to determine some of the coefficients in the ansatz~\eqref{eq:ansatz_6pt_MHV} by imposing consistent factorization into lower-point amplitudes. However, note that the amplitude $M_{6}(--+++\,+)$ only contains 2-particle factorization channels, which involve the product of a 3- and a 5-point amplitude. Therefore, this sector depends at most linearly on the 4-point Wilson coefficients, and is not relevant for our purpose of finding \emph{nonlinear} constraints among them.

\subsection{\texorpdfstring{N$^{(1,0)}$MHV}{N(1,0)MHV} Amplitudes}
\label{sec:6pt_N10MHV}

We now turn to the cases that arise when restricting to $\text{SU}(4) \times\text{SU}(4)$ R-symmetry, starting with the N$^{(1,0)}$MHV. As shown in~\cite{Elvang:2010xn}, the 6-point N$^{(1,0)}$MHV superamplitude is given by
\begin{align}
    & \, \mathcal{M}_6^{\text{N}^{(1,0)}\text{MHV}}(123456) = \frac{\delta^{(16)} \big( \widetilde{Q}\big)}{\<56\>^8} \Bigg[ Y_{1111} \,  
   M_6(z^{1234}+++-\,-) + Y_{(1112)} \, M_6(\chi^{123}\psi^4++--) \nonumber \\
   & \, + Y_{(1122)} \,  
   M_6(v^{12}v^{34}++--) + Y_{(1222)} \,  
   M_6(\psi^1 \chi^{234}++--) + Y_{2222}\,  
   M_6(+\,z^{1234}++--) \Bigg].
\end{align}
In its current form, to determine the superamplitude we would have to construct bottom-up five different EFT component amplitudes. Instead, we can use a different basis, consisting of a single component amplitude and its permutations. For instance, we can project the amplitude $M_6(z^{1234}+++-\,-)$ and four independent permutations, and re-express the superamplitude as\footnote{Note that the naive basis choice, in which the fifth basis element is given by $M_6(+++-z^{1234}-)$, yields a non-invertible change of basis matrix, signaling that these 5 permutations are not totally independent and do not form a basis.}
\begin{align}
    &\, \mathcal{M}_6^{\text{N}^{(1,0)}\text{MHV}}(123456) = C_1 \,  
   M_6(z^{1234}+++-\,-) +  C_2 \,  
   M_6(+\,z^{1234}++--) \nonumber \\[0.2cm]
   &\qquad \qquad +  C_3 \,  
   M_6(++z^{1234}+-\,-) +  C_4 \,  
   M_6(+++\,z^{1234}-\,-) +  C_5 \,  
   M_6(z^{1234}++-+\,-) \,.
\end{align}
The $C_i$ are polynomials of degree 20 in the Grassmann variables, and involve complicated expressions in terms of spinor-helicity variables. With this new basis, one can simply construct $M_6(z^{1234}+++-\,-)$ bottom-up in the EFT with an ansatz analogous to~\eqref{eq:ansatz_6pt_MHV}, and calculate the remaining amplitudes using permutations of the momenta. Nevertheless, this amplitude only involves 2-particle factorizations into 3-point and 5-point N$^{(1,0)}$MHV amplitudes. Thus, it does not depend on the 4-point Wilson coefficients and does not produce constraints among them.

\subsection{\texorpdfstring{N$^{(2,0)}$MHV}{N(2,0)MHV} Amplitudes}
\label{sec:6pt_N20MHV}

Lastly, we briefly consider the N$^{(2,0)}$MHV sector. In this case, the superamplitude can be written in terms of a single component amplitude as~\cite{Elvang:2010xn}
\begin{equation}
    \mathcal{M}_6^{\text{N}^{(2,0)}\text{MHV}}(123456) = \frac{\delta^{(16)} \big( \widetilde{Q}\big)}{\<56\>^8} \, Y_{1111} Y_{2222} \,  
   M_6(z^{1234}z^{1234}++--) \,.
\end{equation}
Just like for the 5-point N$^{(1,0)}$MHV sector from Section~\ref{sec:5pt_amps}, the component amplitude is completely local, as it does not have any possible factorization channels. Thus, given e.g. that
\begin{equation}
    M_6(z^{1234} z^{1234} - + - +) = \frac{\<35\>^4 [46]^4}{\<56\>^4 [34]^4} \, M_6(z^{1234}z^{1234} ++--)\,,
\end{equation}
we deduce that an appropriate ansatz is
\begin{equation}
    M_6(z^{1234}z^{1234}++--) = \ab{56}^4 \sb{34}^4 \, \left( \widetilde{V}_{6}(123456)+\sum_{i=1}^5\epsilon_i\,\widetilde{Q}_{6,i}(123456) \right).
\end{equation}
In this expression, $\widetilde{V}_{6}$ and $\widetilde{Q}_{6,i}$ denote polynomials in the 6-point Mandelstam variables~\eqref{eq:6pt_Mandelstams}, while $\epsilon_i$ denotes a choice of Levi–Civita contraction from~\eqref{eq:6pt_LC}. The arbitrary coefficients in the polynomials correspond to 6-point Wilson coefficients, and are not related (based on the analysis up to 6 points) to the 4-point coefficients. Thus, this sector is also irrelevant for the purpose of finding relations among the 4-point local terms.

\section{Counting \texorpdfstring{$S_2\times S_4$}{S2xS4}-Symmetric Polynomials Using O(4) Characters}
\label{app:counting}

In this appendix, we explain how to adapt character technology from \cite{Chowdhury:2019kaq} (section 4) to count parity-even polynomials in 6-point Mandelstam invariants with $S_2\times S_4$ symmetry in four spacetime dimensions; 
similar methods have been used to enumerate local operators  compatible with various symmetries~\cite{Dolan:2007rq,Gray:2008yu,Lehman:2015via,Henning:2015daa}.

The method considers the Hilbert space of a free field in radial quantization, where for us the states consist of powers of a field and its derivatives: $\partial_{\mu_1}\cdots \partial_{\mu_k}\phi$, where each index runs in $1\ldots D=4$. This Hilbert space enjoys an action of the ${\rm SO}(D)\times {\rm GL}(1)$ group of rotations and dilations, and taking the trace of an element of this group defines a character. A generic element may be written as $x_0^\Delta g(x_1,x_2,\ldots,x_{\lfloor D/2\rfloor})$, where $\Delta=\Delta_\phi + n$ for the preceding monomial and for $D=4$ a general rotation is characterized by its four eigenvalues $(x_1,x_1^{-1},x_2,x_2^{-1})$. Note that there is no difference between the ${\rm SO}(4)$ and ${\rm SO}(3,1)$ groups for the present purposes: spacetime signature determines whether the $x_i$'s are phases or real numbers but all formulas below are rational expressions that make sense for complex $x_i$'s.

The first key ingredient is the trace over a single-particle Hilbert space. Accounting for equations of motion, for a scalar field there is a single symmetric traceless tensor at each derivative order \cite{Chowdhury:2019kaq} (see also \cite{FultonHarris}, section 24.2):
\begin{align}
\nonumber
    Z_1(\{x\}) &\equiv {\rm Tr}_{\rm 1-particle}\left[x_0^{\Delta-\Delta_\phi} g(x_i)\right]
\\ 
\nonumber
&=    \sum_{k=0}^\infty x_0^{k}\,
    \frac{ x_1^{k+1}+x_1^{-k-1}-x_2^{k+1}-x_2^{-k-1}}
    {(x_2-x_1)(1-x_1x_2)/(x_1x_2)} 
\qquad \mbox{for {\rm  SO}(4)}
\\ &= \frac{(1-x_0^2)x_1x_2}{(x_1-x_0)(x_2-x_0)(1-x_0x_1)(1-x_0x_2)} \,.
\end{align}
In the top line we subtracted $\Delta_\phi$ for convenience, since our interest is to count derivatives. The second ingredient is to count total derivatives, e.g.~arbitrary powers of $(p_1^\mu+\ldots+ p_n^\mu)$ without any trace constraint, which gives, again restricting to $D=4$ (see \cite{Chowdhury:2019kaq} for the general case):
\begin{equation}
    Z_\partial = \frac{x_1x_2}{(x_1-x_0)(x_2-x_0)(1-x_0x_1)(1-x_0x_2)}\,.
\end{equation}
The number of Lorentz-invariant polynomials of degree $k$ in $n$ on-shell momenta satisfying momentum conservation is then obtained by suitably averaging over the group and taking the coefficient of $x_0^k$:
\begin{equation} \label{Zn symmetric}
    Z_n^{\rm no\,symmetry}(x_0) = \frac{1}{(2\pi i)^2}\oint \frac{d x_1}{x_1^3} \frac{dx_2}{x_2^2} (x_2-x_1)(1-x_1x_2)  \frac{Z_1(x_0,x_1,x_2)^n}{Z_\partial(x_0,x_1,x_2)}\,,
\end{equation}
where the integrals are to be performed over the unit circle with $x_0$ small.
This amounts, in effect, to taking residues at $x_i=x_0$.
As a simple check, for $n=4$ this yields $Z_n^{\rm no\,symmetry}(x_0)=(1-x_0^2)^{-2}=1+2x_0^2+3x_0^4+\ldots$, which correctly counts polynomials in two independent Mandelstam invariants $s$ and $t$.

We need to refine this in two ways.  The first is to project the $n=6$ Hilbert space onto $S_2\times S_4$-symmetric states.
The general method \cite{Chowdhury:2019kaq} is to insert in the trace of a projector given by the sum over all $2!\times 4!$ permutations, using the fact that each length-$m$ cycle in the permutation yields a factor $Z_m(\{x_i\})=Z_1(\{x_i^m\})$ (corresponding to raising the group element to the $m$'th power). For the case of $S_2\times S_4$, this amounts to the substitution in \eqref{Zn symmetric}:
\begin{equation} \label{S2 S4 twist}
   Z_1^6\mapsto \frac12\left(Z_1^2 + Z_2\right)\times \frac{1}{24}
   \left(Z_1^4+6 Z_1^2Z_2+3 Z_2^2 +8 Z_1 Z_3+6 Z_4\right).
\end{equation}
Summing over the residues inside the unit disc (which are now at $x_i=x_0 \zeta$ with $\zeta$ a third or fourth root of unity) we then obtain
\begin{equation}\begin{aligned} \label{Z6 all}
Z_6^{S_2\times S_4}(s) &= \frac{1}{(1-s)(1-s^2)^2(1-s^3)^2(1-s^4)^2(1-s^6)} \Bigl(1+s^2+3s^3+8s^4+17s^5+30s^6+47s^7
\\&\hspace{-12mm}
+72s^8+81s^9+96s^{10}+97s^{11}+86 s^{12}+74 s^{13}+48s^{14}+32 s^{15}+17 s^{16}+8 s^{17}+s^{18}+s^{19}\Bigr)
\\ &= 1 + s + 4 s^2 + 9 s^3 + 24 s^4 + 53 s^5 + 122 s^6+\ldots\,,
\end{aligned}\end{equation}
where we have set $s=x_0^2$. Thus the coefficient of $s^j$ counts the independent $S_2\times S_4$-symmetric polynomials of degree $j$ in Mandelstam invariants, \emph{as well as} parity-odd structures involving the Levi–Civita tensor.

The second refinement is to project out the parity-odd structures.
The idea is to also consider the trace of an element $Pg$ of O$(4)$ with determinant $-1$, which picks up a minus sign for each Levi–Civita tensor.  Up to group conjugation, such an element is characterized by the eigenvalues $(x_1,x_1^{-1},1,-1)$ and it depends on a single continuous eigenvalue $x_1$ for O(4). For the single-particle and total derivative partition functions we now find
\begin{equation}
    Z_1^{P}=\frac{x_1}{(x_1-x_0)(1-x_0x_1)}\,,\qquad Z_\partial^{P}=\frac{x_1}{(1-x_0^2)(x_1-x_0)(1-x_0x_1)}\,.
\end{equation}
The generating function which counts parity-even minus -odd O(4)-invariants in $n$ on-shell momenta satisfying momentum conservation is then a natural generalization of \eqref{Zn symmetric}:
\begin{equation}
    Z_n^{P,\,\rm no\,symmetry}(x_0) = \frac{1}{2\pi i}\oint \frac{d x_1}{x_1} (1-x_1^2) \, \frac{Z_1^{P}(x_0,x_1)^n}{Z^P_\partial(x_0,x_1)}\,.
\end{equation}
For example for $n=4$ this gives the same result $1/(1-x_0^2)^2$ as discussed below \eqref{Zn symmetric}, in agreement with the fact that all four-particle scalar amplitudes are parity-even in $D=4$ (the four-dimensional Levi–Civita tensor being only nonvanishing with $n\geq 5$ external particles).  To count $S_2\times S_4$ symmetric polynomials we apply the same substitution \eqref{S2 S4 twist} to $(Z_1^P)^6$, but noting in this case that $P$ disappears when the power of the generator is even:
\begin{equation}
    Z_m^{P}(x_0,x_1) = \left\{\begin{array}{ll} 
    Z_1^{P}(x_0^m,x_1^m)\,, \quad & m\,{\rm odd}\,,\\
    Z_1(x_0^m,x_1^m,1)\,, \quad & m\,{\rm even}\,.\end{array}\right.
\end{equation}
In this way we obtain
\begin{equation}\begin{aligned}\label{Z6 P}
Z_6^{P,\,S_2\times S_4}(s) &=
\frac{1}{(1-s)(1-s^2)(1-s^3)^2(1-s^4)(1-s^6)} 
\Bigl(1 + 2 s^2 + 3 s^3 + 7 s^4 + 4 s^5 
\\&
\qquad + 12 s^6 + 4 s^7+ 7 s^8 + 4 s^9 + 
 2 s^{10} - s^{11} + s^{12} - s^{13}\Bigr)\,.
\end{aligned}\end{equation}
Finally, we project onto parity-symmetric polynomials by averaging 
\eqref{Z6 all} and \eqref{Z6 P}, which gives the generating function \eqref{S2 x S4 generating function} recorded in the main text.

\section{6-point NMHV Superamplitude with SU(8) R-symmetry}
\label{app:Rsym_SU8}

The 6-point superamplitude with unbroken SU$(8)$ R-symmetry contains two sectors, namely the MHV and NMHV. The MHV sector automatically preserves SU$(8)$ and was studied in Appendix~\ref{app:6ptHelSect}. In this appendix, we thus study the structure of the 6-point NMHV superamplitude in the case of unbroken SU$(8)$ R-symmetry.

It was shown in~\cite{Elvang:2009wd,Elvang:2010xn} that the 6-point NMHV superamplitude with SU$(8)$ R-symmetry can be written in terms of nine basis amplitudes as
\begin{equation}
\label{N8SuperAmplitude}
    {\mathcal M}_6^\text{NMHV}(123456) =
    \sum_{1\leq i\leq j\leq \ldots \leq v\leq 2}\!\!\!\,X_{(ijklpquv)}\,M_6(\{i\,j\,k\,l\,p\,q\,u\,v\}\, {+}{+}{-}{-})\,.
\end{equation}
Here, $X_{(ijklpquv)}$ denotes the sum over inequivalent permutations of $\{i, j, k, l, p, q, u, v \}$, where the $X_{ijklpquv}$ are given by
\begin{equation}
X_{ijklpquv} \;=\; \delta^{(16)}(\tilde{Q}) \, \frac{m_{i34,1} \, m_{j34,2} \cdots m_{v34,8}}{[34]^8 \langle 56 \rangle^8} \,,
\end{equation}
which depend on the Grassmann polynomials $m_{ijk,a}$ from \reef{eq:mijks}. The nine basis amplitudes are given by
\be
 \label{SU8base}
 \begin{split}
\mathbb{M}_1 &\equiv 
M_6(\{11111111\}\, {+}{+}{-}{-})
= M_6({-}{+} {+}{+}{-}{-})\,,
\\
\mathbb{M}_2 &\equiv 
M_6(\{11111112\}\, {+}{+}{-}{-})
= M_6(\psi^{1234567}\psi^{8} {+}{+}{-}{-})\,,
\\
\mathbb{M}_3 &\equiv 
M_6(\{11111122\}\, {+}{+}{-}{-})
= M_6(v^{123456}v^{78} {+}{+}{-}{-})\,,
\\
\vdots
\\
\mathbb{M}_9 &\equiv 
M_6(\{22222222\}\, {+}{+}{-}{-})
= M_6({+}{-} {+}{+}{-}{-})\,.
 \end{split}
\ee
To switch to the $\mathbb{S}_i$ basis, we pick the permutations
\begin{equation}
\label{nineS1perms}
\big\{
123456,\,
132456,\,
142356,\,
162345,\,
345612,\,
563412,\,
354612,\,
423561,\,
463512
\big\}
\end{equation}
of $\mathbb{S}_1$ as the nine scalar basis amplitudes. We project out each of these from the superamplitude~\eqref{N8SuperAmplitude} to get a linear system
\begin{equation}
\label{eq:changeofbasisSU8}
\begin{pmatrix} \mathbb{S}_1 \\ \mathbb{S}_2 \\ \vdots \\ \mathbb{S}_9 \end{pmatrix}
= C_{9 \times 9}
\begin{pmatrix} \mathbb{M}_1 \\ \mathbb{M}_2 \\ \vdots \\ \mathbb{M}_9 \end{pmatrix}.
\end{equation}
Inverting this linear system, 
the superamplitude is given in terms of the scalar basis as
\begin{equation}
\label{M6fromS1toS9}
    {\mathcal M}_6^\text{NMHV}(123456) =  \sum_{i = 1}^{9} C_{i} \,  \mathbb{S}_{i}\,,
\end{equation}
where the $C_{i}$ are spinor-helicity dependent coefficients.

The bottom-up construction of the EFT amplitude $\mathbb{S}_{1}$ is exactly as in Section \ref{sec:6pt_NMHVeft}, since $\mathbb{S}_{1}$ only has 3-particle factorization channels and therefore is built up from products of 4-point amplitudes which are automatically SU$(8)$-invariant.

\section{Comparison to the Closed Superstring Amplitude}
\label{app:comparison_string}

In this appendix, we gather the KLT relations for the 4-, 5- and 6-point closed string amplitudes, which provide their low-energy expansions in terms of open string amplitudes. These results serve as a complementary check of our EFT amplitudes, since they can be directly compared when fixing the Virasoro–Shapiro amplitude as the corresponding UV completion.

As shown in~\cite{Kawai:1985xq}, any $n$-point closed string amplitude $M_n^{\text{str}}$ can be obtained from the open string ones, denoted as $A_n^{\text{str}}$, by using the so-called KLT relations. First, we consider the 4-point amplitude $M^{\text{str}}_4(--+\,+)$, which is simply the Virasoro–Shapiro amplitude for 4-graviton scattering. In this case, it can be obtained through the KLT relation
\begin{equation}
\label{eq:KLT_4pt}
    M_4^{\text{str}}(--+\,+) = \frac{\sin({\alpha' \pi s})}{ \pi} A_4^{\text{str}}[1^{-} 2^{-} 3^{+} 4^{+}] A_4^{\text{str}}[1^{-} 2^{-} 4^{+} 3^{+}]\,,
\end{equation}
where the 4-gluon open superstring Veneziano amplitude is given by~\cite{Veneziano:1968yb}
\be
A_4^{\text{str}}[1^{-} 2^{-} 3^{+} 4^{+}] =
-g^{2} \<12\>^2 [34]^2 \frac{
\Gamma(1-\alpha' s)
\Gamma(1-\alpha'u)}
{su \, \Gamma(1+\alpha't)} \,,
\ee
and $g$ is the Yang–Mills coupling. Using the Euler reflection formula
\begin{equation}
   \frac{\sin (\alpha' \pi s)}{\pi} = \frac{\alpha's}{ \, \Gamma(1-\alpha' s) \, \Gamma(1+\alpha' s)}\,,
\end{equation}
and the replacement $g^4 \alpha'=\kappa^2$, it is straightforward to recover the Virasoro–Shapiro amplitude given in~\eqref{eq:Virasoro-Shapiro}.

At 5 points, the KLT relation becomes~\cite{Kawai:1985xq}
\begin{equation}
\label{eq:KLT_5pt}
M_{5}^{\text{str}}(--+++)
=
-\frac{\sin\!\big(\alpha' \pi s_{12}\big)\,\sin\!\big(\alpha' \pi s_{34}\big)}{\alpha'^{1/2} \pi^2}\,
A_{5}^{\text{str}}[1^{-}2^{-}3^{+}4^{+}5^{+}]\,A_{5}^{\text{str}}[2^{-}1^{-}4^{+}3^{+}5^{+}]
\;+\; (2 \leftrightarrow 3)\,.
\end{equation}
As shown in~\cite{Mafra:2011nv,Mafra:2011nw} (see also~\cite{Stieberger:2006bh,Stieberger:2006te,Broedel:2009nsh,Schlotterer:2012ny,Broedel:2013aza}), one can obtain the low-energy $\alpha'$-expansion of the 5-point open string amplitudes from Yang–Mills amplitudes via
\begin{equation}
\label{eq:example_5pt_open_F}
\begin{pmatrix}
A_5^{\text{str}}[1 2 3 4 5]\\
A_5^{\text{str}}[1 32 4 5]
\end{pmatrix}
=
F_5
\begin{pmatrix}
A_5^{\text{YM}}[1 2 3 4 5]\\
A_5^{\text{YM}}[1 32 4 5]
\end{pmatrix},
\end{equation}
where $F_n$ denotes a matrix. Its low-energy expansion is given by
\begin{align}
F_n = &\, 1 +\zeta_2\, P_{n,2}+\zeta_3\, M_{n,3}+\zeta_2^2\, P_{n,4}+\zeta_2\zeta_3\, P_{n,2}M_{n,3} + \zeta_5\, M_{n,5} + \zeta_2^3\, P_{n,6} + \ldots\,,
\end{align}
where the matrices $P_{n,i}$ and $M_{n,i}$, which depend on the Mandelstam variables, can be found in~\cite{Stieberger_website}. 

Substituting the previous expressions into the KLT formula in~\eqref{eq:KLT_5pt}, we find that the result agrees with the 5-point EFT amplitude from Appendix~\ref{app:5pt_analysis} up to order $\mathcal{O}(s^9)$ upon replacing the corresponding 4-point Wilson coefficients. At this order, we have the first free coefficient in the ansatz, while the first multi-zeta value $\zeta_{3,3,5}$ enters the 5-point closed string amplitude at order $\mathcal{O}(s^{12})$~\cite{Schlotterer:2012ny}.

At 6 points, a simple way to implement the KLT formula is through the equation~\cite{Mizera:2016jhj}
\begin{equation}
\label{eq:KLT_6point}
   M_6^{\text{str}}=-\frac{1}{\alpha' \pi^3 }(\mathbf{A}_6^{\text{str}})^{T} \,S_6 \,\tilde{\mathbf{A}}_6^{\text{str}}\,,
\end{equation}
where $\mathbf{A}_6^{\text{str}}$ corresponds to the vector of permutations of  $A_6^{\text{str}}[1\,\sigma(234)\, 56]$ in canonical order, and $\tilde{\mathbf{A}}^{\text{str}}_6$ is the vector
\begin{equation}
\tilde{\mathbf{A}}_6^{\text{str}} = \begin{pmatrix}
A_6^{\text{str}}[153462]\\
A_6^{\text{str}}[154362]\\
A_6^{\text{str}}[152463]\\
A_6^{\text{str}}[154263]\\
A_6^{\text{str}}[152364] \\
A_6^{\text{str}}[153264]
\end{pmatrix}.
\end{equation}
Similarly to~\eqref{eq:example_5pt_open_F}, one can compute the 6-point open string amplitudes from Yang–Mills amplitudes; see e.g.~Appendix~C in~\cite{Elvang:2026pmc}. The 6-point KLT kernel $S_6$ is given by a $6 \times 6 $ block-diagonal matrix whose first block is given by~\cite{Bjerrum-Bohr:2010pnr,Mizera:2016jhj}
\begin{equation*}
\begin{pmatrix}
\sin(\alpha' \pi s_{12}) \sin(\alpha' \pi s_{35}) \sin(\alpha' \pi s_{45}) & \sin(\alpha' \pi s_{12}) \sin(\alpha' \pi s_{45}) \sin(\alpha' \pi(s_{34} + s_{35}))\\
\sin(\alpha' \pi s_{12}) \sin(\alpha' \pi s_{35}) \sin(\alpha' \pi (s_{34} + s_{45})) &\sin(\alpha' \pi s_{12}) \sin(\alpha' \pi s_{35}) \sin(\alpha' \pi s_{45})
\end{pmatrix}.
\end{equation*}
The remaining entries in the matrix can be obtained by suitable relabelings.

In particular, we used~\eqref{eq:KLT_6point} to calculate the 6-point graviton closed string amplitude $M_6^{\text{str}}(1^+2^-3^-4^-5^+6^+)$. The result matches the 6-point EFT amplitude from Section~\ref{sec:Nonlinear_constraints} up to order $\mathcal{O}(s^8)$, where the first free 6-point local contact term appears in the ansatz. At 6 points, we expect the appearance of the first multi-zeta value $\zeta_{3,3,5}$ at order $\mathcal{O}(s^{12})$ in the closed string amplitude~\cite{Schlotterer:2012ny}.


\bibliography{maxSUSYrefs}
\bibliographystyle{JHEP}

\end{document}